\documentclass{jpp}
\usepackage{graphicx}
\usepackage{epstopdf, epsfig}
\usepackage{hyperref}
\usepackage{natbib}

\shorttitle{Multidimensional Iterative Filtering for investigating plasma turbulence}
\shortauthor{E. Papini et al.}

\title{Multidimensional Iterative Filtering: a new approach for investigating plasma turbulence in numerical simulations.}

\author{Emanuele Papini\aff{1,2}
  \corresp{\email{papini@arcetri.inaf.it}},
  Antonio Cicone\aff{3},
  Mirko Piersanti\aff{4},
  Luca Franci\aff{5,1},
  Petr Hellinger\aff{6},
  Simone Landi\aff{1,2},
 \and Andrea Verdini\aff{1,2}}

\affiliation{\aff{1}Dipartimento di Fisica e Astronomia, Universit\`a degli Studi di Firenze, via G. Sansone 1, Sesto Fiorentino 50019, Italy
\aff{2}INAF, Osservatorio Astrofisico di Arcetri, Largo E. Fermi 5, Firenze 50125, Italy
\aff{3}Dipartimento di Scienza e Alta Tecnologia, Universit\`a degli Studi dell'Insubria, via Valleggio 11, Como 22100, Italy 
\aff{4}INFN, Sezione di Roma Tor Vergata, Roma, Italy
\aff{5}School of Physics and Astronomy, Queen Mary University of London, London E1 4NS, UK,
\aff{6}Astronomical Institute, CAS, Bocni II/1401, Prague CZ-14100, Czech Republic}

\newcommand{\bv}[1]{\boldsymbol {#1}}
\newcommand{\kpj}{k_\perp^{(j)}}

\newcommand{\IMF}[1]{\widehat{#1}}
\begin{document}

\maketitle

\begin{abstract}
Turbulent space and astrophysical plasmas exhibit a complex dynamics, which involves nonlinear coupling across different temporal and spatial scales. There is growing evidence that impulsive events, such as magnetic reconnection instabilities, lead to a spatially localized enhancement of energy dissipation, thus speeding up the energy transfer at small scales. Capturing such a diverse dynamics is challenging. Here, we employ the Multidimensional Iterative Filtering (MIF) method, a novel technique for the analysis of nonstationary multidimensional signals. Unlike other traditional methods (e.g., based on Fourier or wavelet decomposition), MIF does not require any previous assumption on the functional form of the signal to be identified. 
Using MIF, we carry out a multiscale analysis of Hall-magnetohydrodynamic (HMHD) and hybrid particle-in-cell (HPIC) numerical simulations of decaying plasma turbulence. The results assess the ability of MIF to spatially identify and separate the different scales (the MHD inertial range, the sub-ion kinetic, and the dissipation scales) of the plasma dynamics. Furthermore, MIF decomposition allows to detect localized current structures and to characterize their contribution to the statistical and spectral properties of turbulence. Overall, MIF arises as a very promising technique for the study of turbulent plasma environments.
\end{abstract}

\section{Introduction: Turbulence and Intermittency in space plasmas}

Space and astrophysical plasmas are often found in a turbulent state, characterized by a disordered and chaotic dynamics encompassing many different spatiotemporal scales.
A key aspect of turbulence studies concerns unraveling the physical mechanisms responsible for the transfer and dissipation of energy across such scales.
In-situ spacecraft observations of plasma turbulence in the solar wind \citep{2009bruno,2013chen,2015kiyani} and in the Earth’s magnetosheath \citep{2016stawarz, 2017chen}, show that power spectra of magnetic fluctuations exhibit power-law behaviors encompassing several orders of magnitude in frequency.
At large scales, where the plasma can be described as a fluid within the framework of  Magnetohydrodynamics (MHD), magnetic spectra follow a Kolmogorov-like power-law, which denotes the existence of an inertial range where the scale-to-scale energy transfer takes place, without losses, via interactions between the turbulent eddies \citep{1963iroshnikov,1965kraichnan}.
As the ion characteristic scales (i.e., the ion inertial length $d_i$ and/or the ion gyroradius) are reached, multifluid effects (such as, e.g., the appearance of Hall currents) and ion-kinetic effects become important. Below such scales, we observe a transition to a magnetic power spectrum with a steeper slope (with values ranging from -2 to -4, but tipically around -3), until dissipation scales are reached \citep[for a comprehensive review, see][and references therein]{2019verscharen}.


Due to the complexity of the kinetic plasma dynamics,
we still lack a definite explanation for the existence of a turbulent cascade beyond the inertial range. 
Several attempts invoke nonlinear wavelike interactions of dispersive modes \citep[described in terms of, e.g.,  kinetic alfv\'en and/or whistler waves,][]{2008howes,2009scheko}, eventually complemented by kinetic dissipative effects, such as Landau damping \citep{Sulem_al_2016} and other effects associated to the deformation of the particles velocity distribution functions \citep{DelSarto_Pegoraro_2018,2017yang}. 
From the other side, numerical evidence hints that the MHD inertial range extends beyond ion scales, provided that one includes the Hall term in the generalized Ohm's law \citep{2018hellinger}, so that the steeper slope is not caused by any dissipative process.

This scenario gets further complicated by the existence of coherent structures, such as current sheets, which naturally form in turbulent environments. These are related to a phenomenon known as intermittency \citep{1995Frisch,1997marsch}, that is, the occurrence of sudden changes in the magnetic fluctuations which lead to a spatially inhomogeneous energy cascade and dissipation.
Indeed, there is growing evidence that plasma instabilities, such as magnetic reconnection triggered in spatially localized current sheets, enhance magnetic energy dissipation  \citep{2018camporeale}.
This casts some doubts on the interpretation of the turbulent cascade in terms of wavelike modes only, and more in general on models that do not include intermittent effects from coherent structures. 
In this context, theoretical models introducing the concept of reconnection-mediated turbulence, have been proposed \citep[among others,][]{2012boldyrev,2017loureiro,2017mallet,2019landi_arxiv}.

Nowadays, investigating plasma turbulence using direct approaches is becoming more and more feasible. The increasing computational capabilities allow to run direct numerical simulations, retaining the main physics ingredients at microscales 
\citep[e.g.,][]{2011howes,2012servidio,2015wan,2017haggerty}. 
In particular, large high-resolution hybrid Particle-In-Cell (PIC) numerical simulations (using a kinetic description for the ions and modeling the electrons as a fluid), later complemented by Hall-magnetohydrodynamic (HMHD) simulations, were able to reproduce most of the turbulent properties observed in the solar wind \citep{2015franci_a,2015franci_b,2018franci,Franci_al_2018,2019franci,2019papini_turb} and in the Earth magnetosheath \citep{2020franci}. 
Such simulations also showed that the development of a turbulent cascade at sub-ion scales is concurrent to the onset of reconnection events in ion-scale current sheets \citep{2016franci,2017cerri,2017franci}. 
Moreover, spectral properties at the reconnection exhausts consistent with a developed turbulent state were observed in a fully kinetic simulation \citep{2017pucci}.
Finally, recent works \citep{2017franci,2019papini_sohe,2019papini_turb} have quantitatively shown
that current sheets undergoing reconnection in developing turbulence trigger an energy transfer directly from large to small scales, 
and can initiate a turbulent cascade that later establishes a proper inertial range, regardless of the model (MHD, Hall-MHD, or ion-kinetic) employed.


The ability of numerical experiments to reproduce the turbulent plasma properties is encouraging, as it confirms that the physical models employed are the right ones to explain spacecraft observations (in that range of scales and/or in those specific plasma conditions).
Moreover, unlike in situ observations, numerical simulations provide the full spatiotemporal information needed to understand the plasma dynamics. Nevertheless, extracting such information is challenging.
Traditional analysis techniques, based on Fourier or wavelet decomposition, have been successful in describing some statistical properties of turbulence \citep[e.g.,][]{2001bruno,2004chang,2005consolini,2008horbury,2016lion}.
Such methods, however, assume stationarity and/or linearity of the signal to be analyzed. Yet, turbulence is intrinsically nonlinear and nonstationary.

To address these limitations, \citet{1998huang} developed the Empirical Mode Decomposition (EMD), a technique specifically designed for decomposing nonstationary nonlinear one-dimensional signals into a set of Intrinsic Mode Functions (IMF), that oscillate around zero but with varying frequency and amplitude. 
Such decomposition is adaptive, based on the local characteristic scales of the signal, and does not require any assumption on the shape of the signal to be extracted. EMD has proven to be a very powerful tool in many research areas and has recently been used to measure the multifractal properties of the solar wind \citep{2019alberti}. Unfortunately, EMD shown to be unstable in presence of noise, and the Ensemble EMD \citep[EEMD,][]{2009wu} and similar alternative methods, which address this issue, greatly increase the computational costs and lack a rigourous mathematical theory behind them.

As an alternative to (E)EMD and equivalent techniques, algorithms based on Iterative Filtering (IF) have been recently developed \citep{2009lin,2016cicone,2020cicone}. Unlike EMD, they give a convergent solution for any square-integrable ($L^2$) signal, also in presence of noise.
IF methods have already been successfully employed in the analysis of time-series from geomagnetic measurements \citep{2018piersanti,2018bertello,2019spogli}.
Multidimensional Iterative Filtering (MIF) generalizes IF to high-dimensional signals, and represents the fastest and more robust adaptive multidimensional decomposition technique currently available in the literature.
It outperforms other methods in terms of computational costs and, at the same time, it retains all the convergence properties of the one-dimensional IF algorithms \citep[for more details, see][]{2017Cicone,2020cicone_fft,2020cicone}.

In this work, we carry out the first multiscale analysis of numerical simulations of plasma turbulence by means of MIF decomposition.
We focus on two numerical datasets, obtained from one HMHD and one HPIC simulation respectively.
 
Our results demonstrate the ability of MIF to: (i) separate the different turbulent regimes (the Energy injection scales, the MHD inertial range, the sub-ion kinetic regime, and the dissipation scales) while retaining the information about the magnetic field  spatial configuration, (ii) disentangle the morphological and physical features of magnetically reconnecting current sheets, and (iii)  quantify the statistical properties of turbulence.

\section{Numerical simulations of plasma turbulence}
The datasets used in this work were produced by two high-resolution numerical simulations of plasma turbulence, thoroughly caracterized in \citet{2019papini_turb}. a Hall-MHD and a Hybrid-PIC simulation. 

\subsection{The Hall-MHD model}
The HMHD model takes into account two-fluids effects that describe the separate dynamics of ions and electrons at sub-ion scales. Different HMHD models can introduce several levels of complexity, depending on whether they retain a description for the pressure tensors and/or for electron inertia effects \citep[see, e.g.,][]{2001shay}. Here we use a model that consists of the nonlinear viscous-resistive MHD equations, modified only by the presence of the Hall term in the induction equation. This is done by substituting the fluid velocity $\boldsymbol{u}$ with the electron velocity $\boldsymbol{u}_e = \boldsymbol{u} - \boldsymbol{J}/ e n_e$. In their adimensionalized form, the HMHD equations take the form
\begin{eqnarray}
 \p_t \rho + \bnabla\cdot{(\rho \boldsymbol{u})} & = & ~0,
  \label{eq:continuity}
  \\
 \rho\left (\p_{t} + \boldsymbol{u}\cdot\bnabla\right) \boldsymbol{u} & = & -\bnabla P + (\bnabla\times{\boldsymbol{B}})\times\boldsymbol{B} + \nu \left [ \nabla^2\boldsymbol{u} + \frac{1}{3}\bnabla(\bnabla \cdot {\boldsymbol{u}}) \right],\\
  \left (\p_{t} + \boldsymbol{u}\cdot\bnabla \right) T & = &
  ~(\Gamma  -  1) \! \left \{ - (\bnabla\cdot{\boldsymbol{u}})T 
  \! + \!  \eta \frac{|\bnabla\times\boldsymbol{B}|^2}{\rho} \right .\nonumber \\
   && \left . +  \frac{\nu}{\rho} \left [ (\bnabla\times{\boldsymbol{u}})^2  +\frac{4}{3} (\bnabla\cdot{\boldsymbol{u}})^2\right ] \right \}
  ,   \\
   \p_t{\boldsymbol{B}} & = & ~\nabla\!\times\!\left ( \boldsymbol{u} \times \boldsymbol{B}\right ) \! + \! \eta \nabla^2 \boldsymbol{B} -  \eta_\mathrm{H}\nabla\!\times\! \frac{(\nabla\!\times\!\boldsymbol{B})\!\times\!\boldsymbol{B}}{\rho},
  \label{eq:induction_hall_adi}
\end{eqnarray}
where $\Gamma=5/3$ is the adiabatic index and $\{\rho,\bv{u},\bv{B},T,P\}$ are a function of time and space and denote the usual variables. The equation of state $P=\rho T$ relates the gas pressure to the other two thermodynamic variables. All quantities are renormalized with respect to a characteristic length $L=d_i$, a magnetic field amplitude $B_0$, a plasma density $\rho_0$, an Alfv\'en velocity $c_A = B_0/\sqrt{4\pi\rho_0} = \Omega_i d_i$, a pressure $P_0=\rho_0 c_A^2$, and a plasma temperature $T_0 = (k_B/m_i) P_0/\rho_0$. Moreover $\Omega_i = e B_0 / m_i c$ is the ion-cyclotron angular frequency and $m_i$ is the mass of the ions. With this normalization, the (adimensional) magnetic resistivity $\eta$ is in units of $d_i c_A$ and the Hall coefficient $\eta_H=d_i/L$ is equal to 1. 

The equations (\ref{eq:continuity}-\ref{eq:induction_hall_adi}) are numerically solved by using a pseudospectral code we developed, already employed for studies of magnetic reconnection \citep{2015landi,2018papini,2019papini_rec} and plasma turbulence \citep{2019papini_sohe,2019papini_turb}.
We consider a two-dimensional $(x,y)$ periodic domain and use Fourier decomposition to calculate the spatial derivatives. In Fourier space we also filter according to the $2/3$ Orszag rule, to avoid aliasing of the nonlinear terms. For the temporal evolution of $\{\rho,\bv{u},\bv{B},T\}$ we use a 3rd-order Runge-Kutta scheme.
\subsection{The Hybrid-PIC model}

The second dataset was produced by using the Lagrangian HPIC code CAMELIA \citep[Current Advance Method Et cycLIc leApfrog, ][]{1994matthews,2018franci}. 
In CAMELIA, the ions are modeled as macroparticles that correspond to statistically-representative portions of the distribution function in the phase space. The plasma charge is neutralized by a massless and isothermal electron fluid. 
The system is governed by the Vlasov-Maxwell equations. Electron inertia effects and the displacement current in the Maxwell's equations are neglected. Therefore, only macroparticle's position and velocity inside each grid cell, as well as  magnetic fields defined at the cell nodes, need to be evolved in time. All other quantities and moments are functions of the above quantities, including the electric field \citep{1994matthews}.

Among many applications, CAMELIA  has been employed for numerical studies of plasma turbulence \citep[e.g.][]{2015franci_a,2015franci_b,2016franci,2016bFranci,2017franci}. It reproduced many of the spectral properties observed in the solar wind \citep{2018franci} and in the Earth magnetosheath \citep{2019franci} (we refer the reader to \citet{2018franci} for further details and applications).

\subsection{Numerical setup}

Apart from few parameters, the HMHD and the HPIC simulations employ the same setup.
We consider a 2D box of size $L_x \times L_y = 256~d_i \times 256~d_i$ and a grid resolution of $\Delta x = \Delta y = d_i/8$, corresponding to $2048^2$ points.
The system is initialized with a constant mean magnetic field $\bv{B}_0 = B_{0} \bv{e}_z$ out of the plane, along the $z$ direction (that we will refer to as the parallel direction). The $xy$-plane (i.e. the perpendicular plane) is filled with freely-decaying random Alfv\'enic-like sinusoidal fluctuations. These are characterized by a root-mean-square amplitude $b_\mathrm{rms} = B_\mathrm{rms}/B_0 \simeq 0.24$, and wavenumbers spanning from the smallest nonzero value contained in the box up to the  injection scale $\ell_\mathrm{inj}=2\pi/k_\perp^\mathrm{inj}$, such that  $k_\perp^\mathrm{inj} d_i \simeq 0.28$, with  $k_\perp = \sqrt{k_x^2+k_y^2}$. 
In the HPIC simulation, we set the ion and electron plasma beta to $\beta_i=\beta_e=1$, while the magnetic resistivity has the value $\eta=5\times 10^{-4}$. The HMHD simulation has a (total) plasma $\beta=\beta_i+\beta_e=2$, and a resistivity and a viscosity $\eta=\nu =10^{-3}$.

\subsection{Datasets of fully developed turbulence}
\label{sec:turb_datasets}
\begin{figure}
 \includegraphics[width=\textwidth]{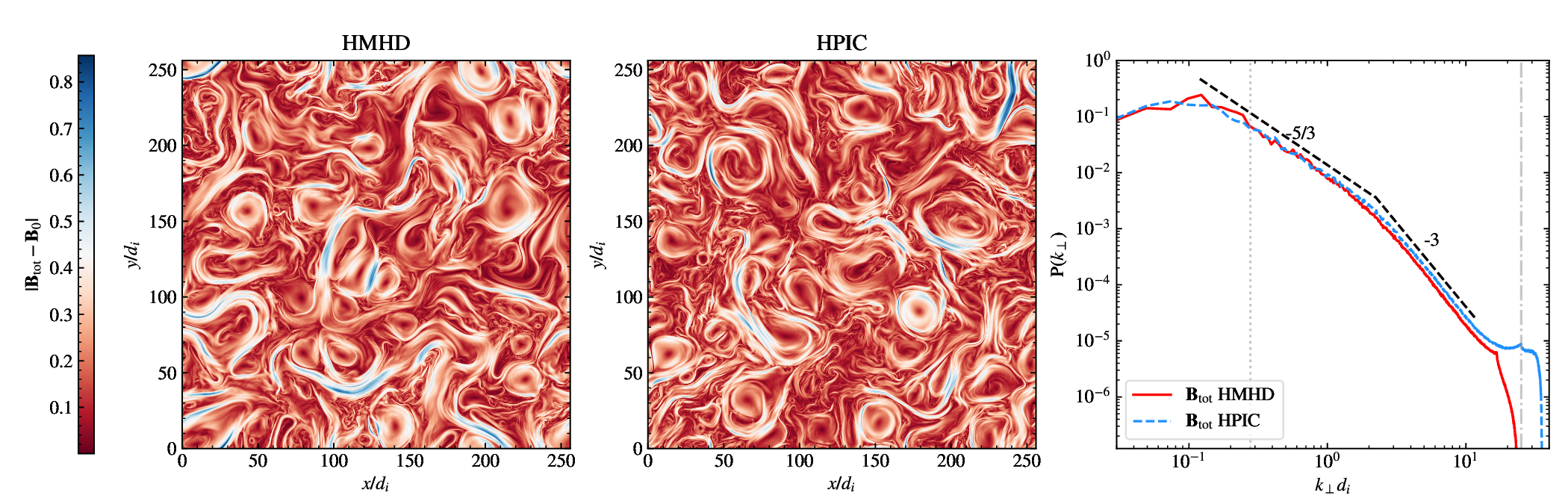}
 \caption{Coloured contours of the amplitude of the total magnetic field fluctuations  $|\bv{B}_{tot} - \bv{B}_0|$, at the maximum of the turbulent activity for the HMHD run at $t=165~\tau_A$ (left panel) and for the HPIC run at $t= 200~\tau_A$ (middle panel). The corresponding isotropized power spectra \citep{2019papini_turb} (solid red curve for the HMHD dataset and dashed blue curve for the HPIC run) are shown on the right panel. Vertical dotted and dot-dashed lines denote the injection wavenumber $k_\perp^\mathrm{inj}$ and the Nyquist wavenumber respectively.}
\label{fig:datasets}
 \end{figure}
In both simulations, the initial Alfv\'enic fluctuations quickly evolve to form coherent structures, namely vortices and, in between them, current sheets. The latter get disrupted by magnetic reconnection and release small-scale plasmoids which feed back to their turbulent sorrounding. 
The subsequent evolution is characterized by the formation and disruption of many other current sheets. Concurrently, a turbulent cascade develops at large scales, until a quasi-stationary state, shown in Figure \ref{fig:datasets}, is reached at $t=165~\tau_A$ and at $t=200~\tau_A$ for the HMHD and the HPIC run respectively. At those times, the power spectrum of the total magnetic fluctuations (right panel of Fig. \ref{fig:datasets}) has a clear multiscale behavior. At scales larger than the injection scale ($k_\perp < k_\perp^\mathrm{inj}$) we have the reservoir of energy that fuels the cascade. At fluid MHD scales ($k_\perp^\mathrm{inj} <k_\perp \lesssim 2/d_i$) a Kolmogorov-like power law of spectral index $-5/3$ is present, which then transitions to a slope of $-3$ at sub-ion kinetic scales at about $k_\perp^\mathrm{break} \sim 2/d_i$ (the so called spectral break). Finally, at $k_\perp > k_\perp^\mathrm{diss} \simeq 12/d_i$ we reach the dissipation scales. The physical and statistical properties of these four regimes are quite different, due to the diversity of the underlying physical mechanims acting at those scales. For instance, sub-ion scales are characterized by increasing levels of intermittency generated by the presence of thin localized current structures where dissipation is enhanced. We refer the reader to In the next sections, we will show how MIF methods can correctly separate these regimes.

\section{Multidimensional Iterative Filtering}
\label{sec:MIF}
We now introduce the Multidimensional Iterative Filtering technique. 
For more details about MIF methods, as well as applications and examples, we remind the reader to \citet{2017Cicone}.

Given a (multidimensional) signal, $f(\bv{r})$ with $\bv{r} \epsilon \mathbb{R}^k$, MIF decomposes it into a finite number $N$ of (locally almost orthogonal) simple oscillating components $\IMF{f}$ called Intrinsic Mode Functions (IMF)

\begin{equation}
 f(\bv{r}) = \sum_{j=1}^N \IMF{f}_j(\bv{r}) + r_{f,N}(\bv{r}),
\end{equation}
where $r_{f,N}$ is the residual of the decomposition (ideally, a trend signal).
Each $\IMF{f}_j$ is the result of an iterative procedure that uses a low-pass filter to extract the moving average of the signal at a given scale $\lambda_j$, so to isolate a fluctuating component whose average frequency $\nu_j\simeq 1/\lambda_j$ is well behaved.  $\lambda_j$ is different for each IMF and increasing with $j$. Therefore, IMFs with increasing $j$ will contain larger (smaller) scales (frequencies).
 
\begin{figure}
\begin{center}
 \includegraphics[width=0.6\textwidth]{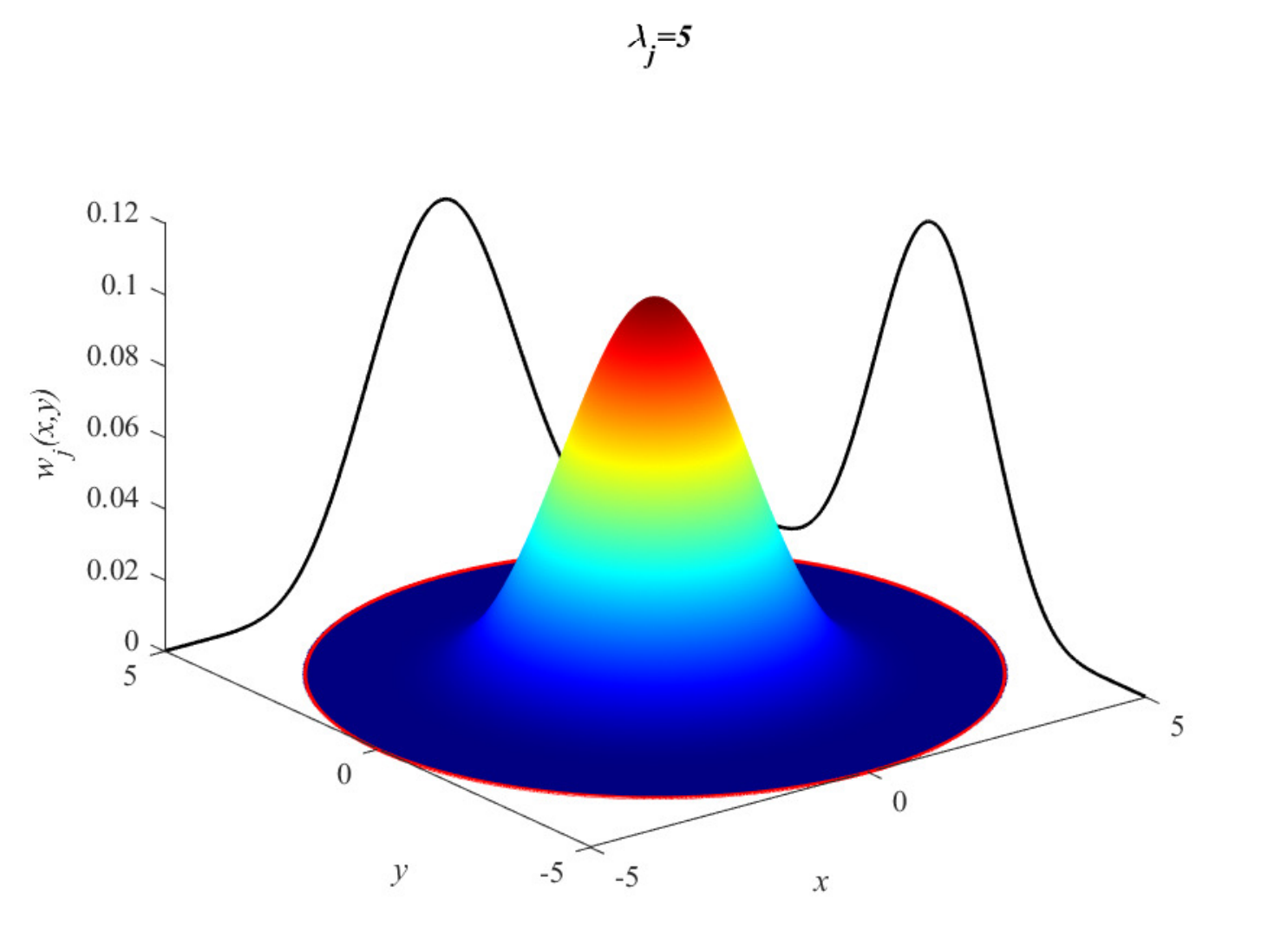}
\end{center}
 \caption{Kernel function $w_j(x,y)$ for a MIF decomposition of a two-dimensional field. The red circle of radius $\lambda_j=5$ denotes the boundary of $\Omega(\lambda_j)$.}

\label{fig:2dwindow_function}
 \end{figure}

We first specify the low-pass filter operator
\begin{equation}
 \mathcal{L}_j [s(\bv{r})] = \int_{\Omega(\lambda_j)} s(\bv{r} + \bv{t}) w_j(\bv{t}) d\bv{t}
\end{equation}
that acts on a $L^2$ signal $s(\bv{r})$. Here $w_j(\bv{r}) \,\epsilon\, \Omega(\lambda_j)$ is the kernel function associated to the filter, and $\Omega(\lambda_j) \subset \mathbb{R}^k$ is the spherical support of $w_j$ with radius $\lambda_j$ (e.g., a circle in $\mathbb{R}^2$).
In this work we use a two-dimensional isotropic kernel function (see Figure \ref{fig:2dwindow_function}), with a Fokker-Planck radial profile and periodical boundary conditions \citep{2016cicone,2017Cicone,2020cicone_boundary}. 

Let us now define $S_{1,0}(\bv{r}) = f(\bv{r})$ and introduce the fluctuating function $S_{1,1}(\bv{r}) = S_{1,0} (\bv{r})- \mathcal{L}_1[S_{1,0}(\bv{r})]$. The scale $\lambda_1$ of $\mathcal{L}_1$ is chosen such that its lenght is comparable to the maximum frequency contained in $S_{1,0}$ \citep[for more details on the choice of $\lambda_1$, see][]{2009lin,2016cicone}.
Through iteration, one can calculate $S_{1,n}(\bv{r}) = S_{1,n-1}(\bv{r}) - \mathcal{L}_1[S_{1,n-1}(\bv{r})]$.
The first IMF is obtained in the limit of infinite iterations
\begin{equation}
 \IMF{f}_1(\bv{r}) = \lim_{n\rightarrow\infty} S_{1,n}(\bv{r}),
\end{equation}
and the residual signal is
\begin{equation}
 r_{f,1} (\bv{r})= f(\bv{r}) - \IMF{f}_1(\bv{r}).
\end{equation}
The second $\IMF{f}$ can be calculated by defining $S_{2,0}(\bv{r}) = r_1(\bv{r})$ and repeating the above procedure but using a kernel function with a larger radius $\lambda_2 > \lambda_1$, whose value is chosen based on $S_{2,0}$. 

The $j$-th IMF is given by
\begin{equation}
 \IMF{f}_j(\bv{r}) = \lim_{n\rightarrow\infty} S_{j,n}(\bv{r}), 
\end{equation}
with 
\begin{equation}
 S_{j,n}(\bv{r}) = S_{j,n-1}(\bv{r}) - \mathcal{L}_j[S_{j,n-1}(\bv{r})],
 \quad S_{j,0}(\bv{r}) = r_{f,j}(\bv{r})
\end{equation}
and 
\begin{equation}
 \lambda_j > \lambda_{j-1} > \dots > \lambda_1.
\end{equation}
The decomposition ideally ends when the residual $r_{f,N}(\bv{r})$ is a trend signal, that is, $r_{f,N}$ contains no local extrema. In practice this is achieved by requiring that $r_{f,N}(\bv{r})$ contains less than a given number of local extrema. In the following, we drop the $N$ subscript to simplify notation and write the residual of a signal $f$ as $r_f$.

\section{Multiscale analysis of fully developed Turbulence}
\begin{figure}
 \includegraphics[width=\textwidth]{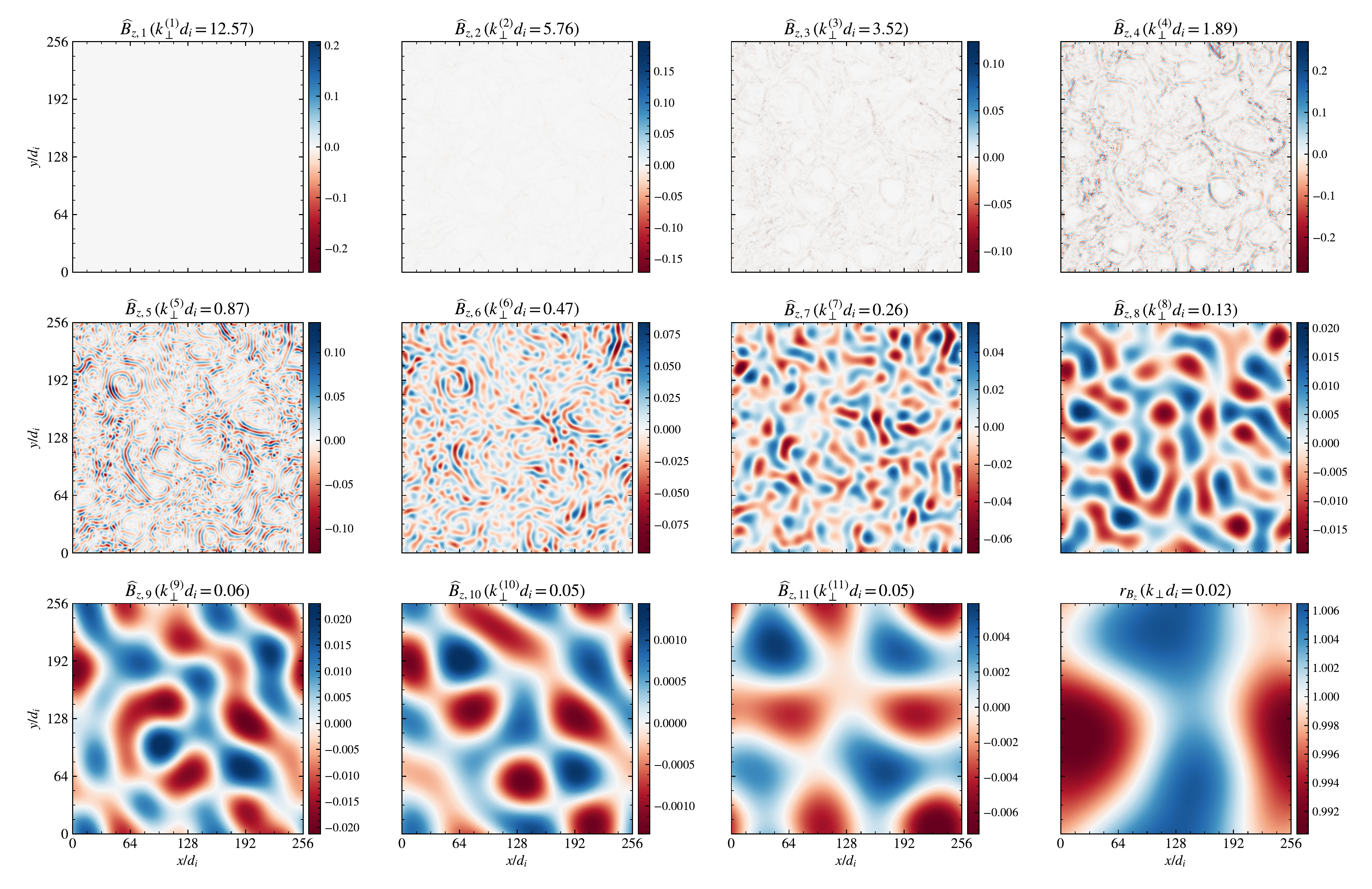}
 \caption{Colored contours of the IMFs $\IMF{B}_{z,j}$ resulting from the MIF decomposition of the out-of-plane magnetic field fluctuations $B_z$  of the HPIC dataset (from small
to large scales going from left to right and from top to bottom). 
  The residual $r_{B_z}$ (bottom right panel) contains the largest scale field fluctuations and the mean field $B_0$. For each IMF $\IMF{B}_{z,j}$, its average spatial wavenumber $\kpj = 2\pi \nu_j$ (see Section \ref{sec:MIF}) is reported.}
 \label{fig:imfs_bz}
\end{figure}

We now describe the multiscale analysis performed on the turbulent magnetic fields from the two numerical datasets described in Section \ref{sec:turb_datasets} and shown in Fig. \ref{fig:datasets}.
In Figure \ref{fig:imfs_bz}, we report the MIF decomposition of the out-of-plane component of the magnetic field, $B_z$, from the HPIC run. Eleven IMFs $\{\IMF{B}_{z,j}$ with $j=1,...,11\}$ have been extracted. The residual $r_{B_z}(x,y)$ is shown on the bottom right panel. The dissipation and the ion kinetic scales (high spatial frequencies, $k_\perp d_i > 2$) are captured by the first four IMFs and are characterized by well localized structures. Going to larger scales ($k_\perp d_i < 2$), these features become more homogenously distributed. That happens, for instance, in the MHD inertial range, captured by $\IMF{B}_{z,5},\IMF{B}_{z,6}$, and $\IMF{B}_{z,7}$. Finally, the largest scales (above the injection) are contained in the last IMFs and in the residual, which also retain the mean field $B_0$. Similar results (not shown here) are obtained for $B_x$ and $ B_y$, as well as for the MIF decomposition of the  HMHD simulation.
Unlike wavelets and Fourier modes, the IMFs are only locally almost orthogonal. Therefore, it may be useful to assess the degree of orthogonality. The orthogonality of a set $\{\IMF{f}_i(x,y)\}$ is given by the (symmetric) orthogonality matrix $\mathbf{M}$, with elements
\begin{equation}
 \mathrm{M}_{ij}=\langle \IMF{f}_i,\IMF{f}_j\rangle = \frac{1}{ \|\IMF{f}_i\|\cdot\|\IMF{f}_j\|}  \left |\int_0^{L_x} \int_0^{L_y} \IMF{f}_i (x,y) \IMF{f}_j (x,y)\, \mathrm{d}x \mathrm{d}y  \right |,
 \label{eq:Mij}
\end{equation}
where 
\begin{equation}
 \|\IMF{f}_i\| = \left ( \int_0^{L_x} \int_0^{L_y} \IMF{f}_i (x,y)^2 \,\mathrm{d}x \mathrm{d}y  \right )^{1/2}.
\end{equation}
The set is orthogonal if $\mathrm{M}_{ij}  = \delta_{ij}$ for each $i,j$.
\begin{figure}
\begin{center}
 \includegraphics[width=.5\textwidth]{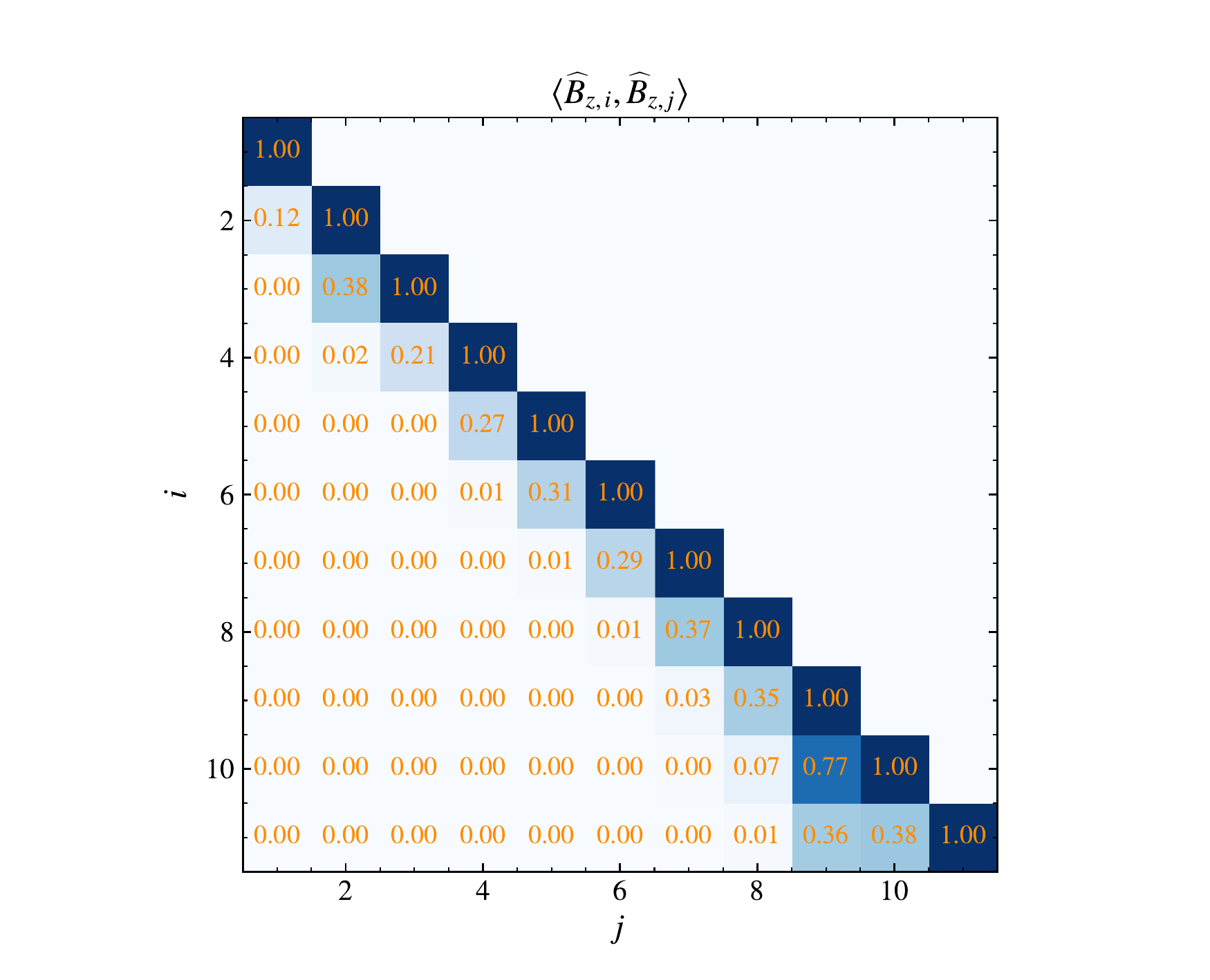}
\end{center}
 \caption{IMF Orthogonality matrix $\mathbf{M}$ of the out-of-plane magnetic field fluctuations, as given by Eq.~(\ref{eq:Mij}).  Only the lower triangle is shown.}
 \label{fig:orto_bz}
\end{figure}
Figure \ref{fig:orto_bz} shows the orthogonality matrix of the IMFs of the out-of-plane magnetic field fluctuations (see Fig. \ref{fig:imfs_bz}).
As espected, the set is not orthogonal, since for neighbor IMFS the lower diagonal (indices $(i+1,i)$) of the orthogonality matrix reaches a maximum value of $0.77$ and a mean of $0.34$. However, for second neighbors ($(i+2,i)$ pairs) the values drop to less than $0.08$ (except for one point).

The components of the magnetic fluctuations at large injection, inertial range/MHD, ion kinetic, and at  dissipation scales are further separated by regrouping the IMFs in four aggregated (vector) IMFs $\{ \IMF{\boldsymbol{B}}_\mathrm{inj}, \IMF{\boldsymbol{B}}_\mathrm{MHD}, \IMF{\boldsymbol{B}}_\mathrm{kin}, \IMF{\boldsymbol{B}}_\mathrm{diss} \}$. Their components (for, e.g., $B_x$) are defined as 
\begin{eqnarray}
 \IMF{B}_{x,\mathrm{inj}} =\sum_{\kpj=0}^{k_\perp^\mathrm{inj}} \IMF{B}_{x,j} + r_{B_x},\quad
 \IMF{B}_{x,\mathrm{MHD}} = \sum_{\kpj>k_\perp^\mathrm{inj}}^{2/d_i} \IMF{B}_{x,j}\nonumber\\
 \IMF{B}_{x,\mathrm{kin}} = \sum_{\kpj>2/d_i}^{k_\perp^\mathrm{diss}} \IMF{B}_{x,j},\quad
 \IMF{B}_{x,\mathrm{diss}} = \sum_{\kpj>k_\perp^\mathrm{diss}} \IMF{B}_{x,j}
 \label{eq:aggregated_imfs}
\end{eqnarray}
where $\kpj$ is the average spatial wavenumber of $\IMF{B}_{x,j}$ 
and, for each aggregated IMF, the sum is performed over the range of scales of interest.
\begin{figure}
  \includegraphics[width=\textwidth]{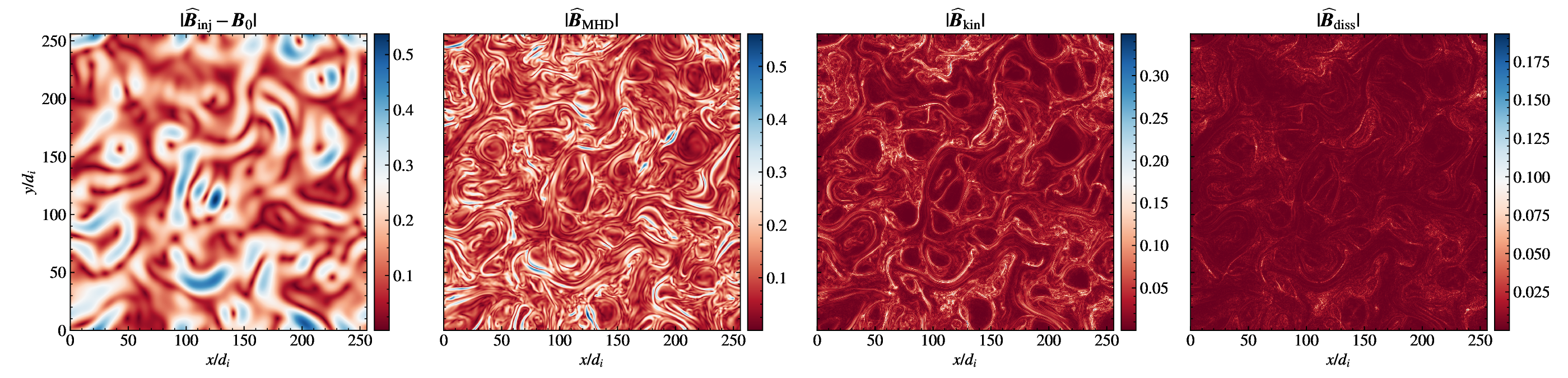}
  \includegraphics[width=\textwidth]{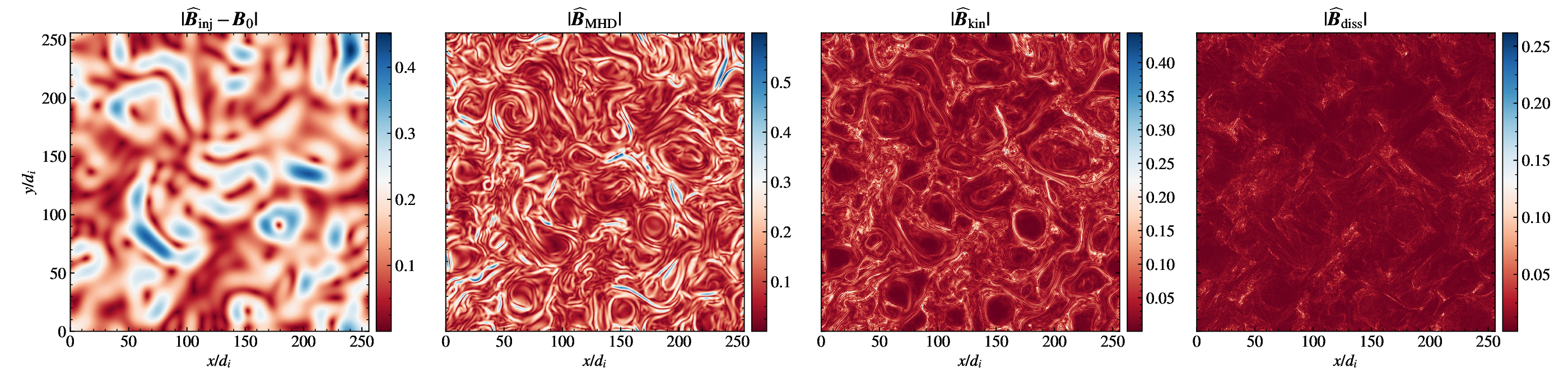}
  \caption{From left to right: amplitude of the aggregated IMFs (see Eq. (\ref{eq:aggregated_imfs})) of magnetic fluctuations at the injection ($|\IMF{\bv B}_{\mathrm{inj}}-\boldsymbol{B}_0|$),
 fluid ($|\IMF{\bv B}_{\mathrm{MHD}}|$), ion kinetic ($|\IMF{\bv B}_{\mathrm{kin}}|$), and
dissipation ($|\IMF{\bv B}_{\mathrm{diss}}|$) scales.
  Top and bottom panels refer to the HMHD and to the HPIC run, respectively. The corresponding power spectra are shown in Fig. \ref{fig:imf_spectra}.
}
\label{fig:aggregated_imfs_B}
\end{figure}
\begin{figure}
 \includegraphics[width=.5\textwidth]{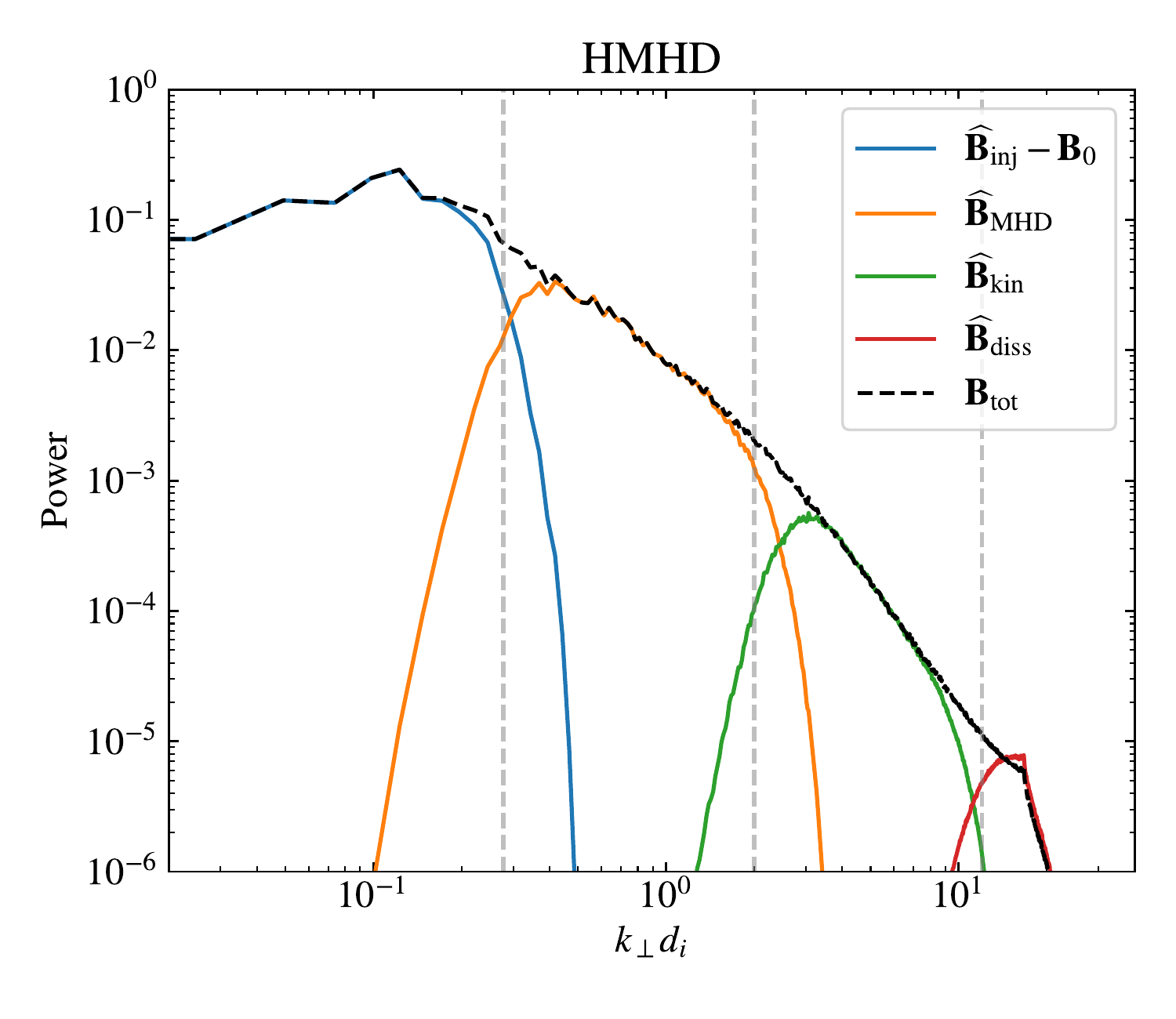}
 \includegraphics[width=.5\textwidth]{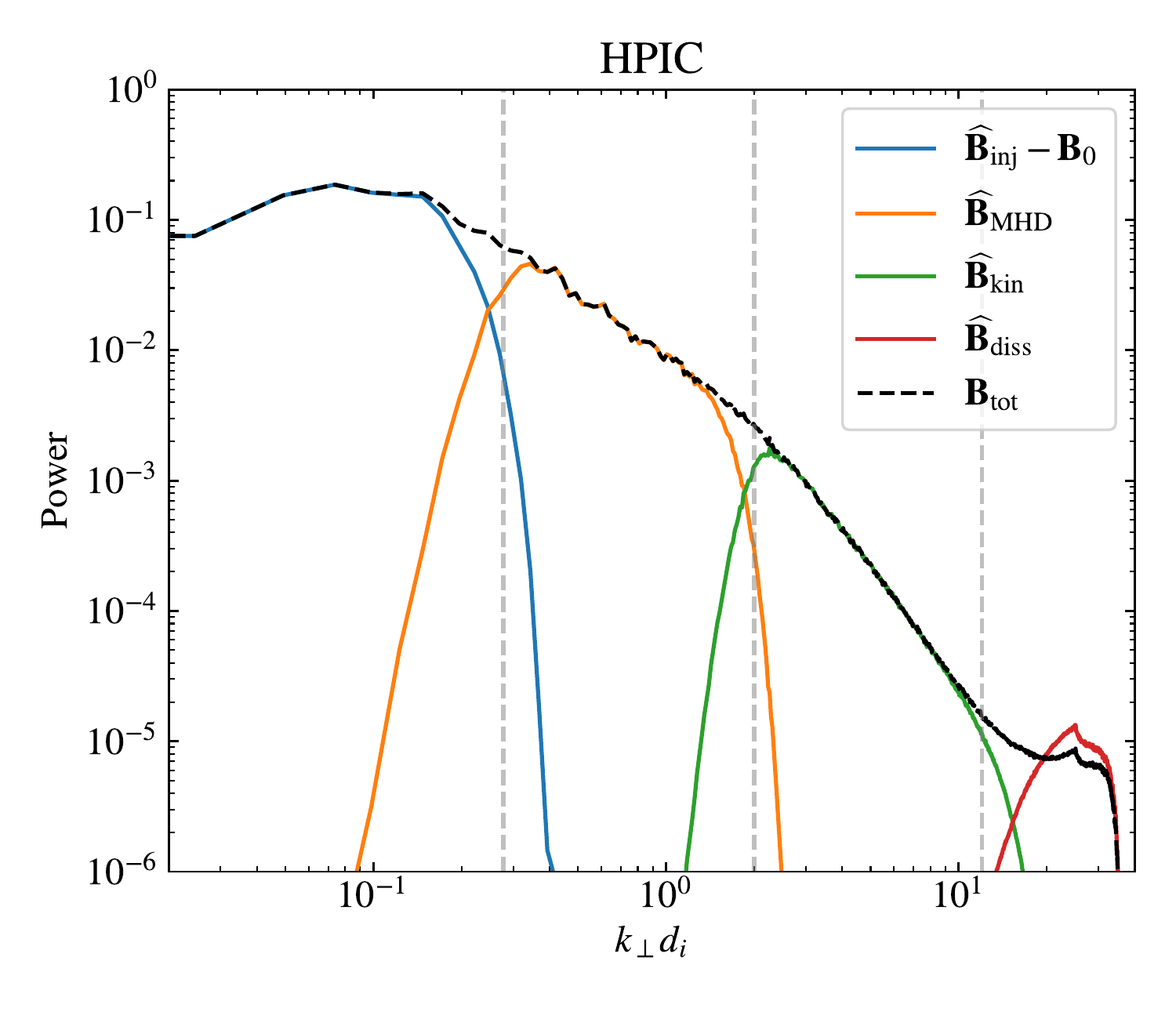}
 \caption{Isotropized 1D power spectra of the total magnetic field fluctuations $\boldsymbol{B}$ (black dashes) and of the injection scales (blue), the MHD scales (orange), the ion kinetic scales (green), and the dissipation scales (red) agregated IMFs, for the HMHD run (left) and for the HPIC run (right). Vertical dashed lines denote $k_\perp^\mathrm{inj}$, $k_\perp d_i=2$, and $k_\perp d_i =12$, i.e., the three wavenumbers that approximately separate the four regimes.
}
\label{fig:imf_spectra}
\end{figure}
The amplitude of the agregated IMFs (Fig. \ref{fig:aggregated_imfs_B}) reveals that the HPIC and the HMHD run are morfologically equivalent, characterized by homogenously distributed features at large scales. As the scales decrease,  such features become  more and more localized, self-organizing in a filamented network  at the edge of the turbulent eddies, where magnetic dissipation is enhanced.  The isotropized power spectra of the agregated IMFs are shown in Fig. \ref{fig:imf_spectra}. 
The MHD inertial range, the kinetic range, and the dissipation scales, as well as the injection scales range, are well separated also in Fourier space.
This further confirms the ability of the MIF decomposition to succesfully isolate the four different regimes, while retaining the full spatially local information of the fields.

As a final remark, the agregated IMFs have an increased orthogonality, as the maximum value out of the diagonal in their ortoghonality matrix is about 0.10.

\subsection{Current structures and intermittency}
\label{sec:intermittency}

Intermittency in plasma turbulence is related to the dynamics of current sheets and localized coherent structures, since they break the self-similarity of the system. Usually, such structures are found in a sort of multifractal configuration, such as the filamented network we observed in Figure \ref{fig:aggregated_imfs_B}.
With MIF we can easily isolate these features.
\begin{figure}
  \includegraphics[width=\textwidth,trim={0.6cm 0 0.5cm 0cm},clip]{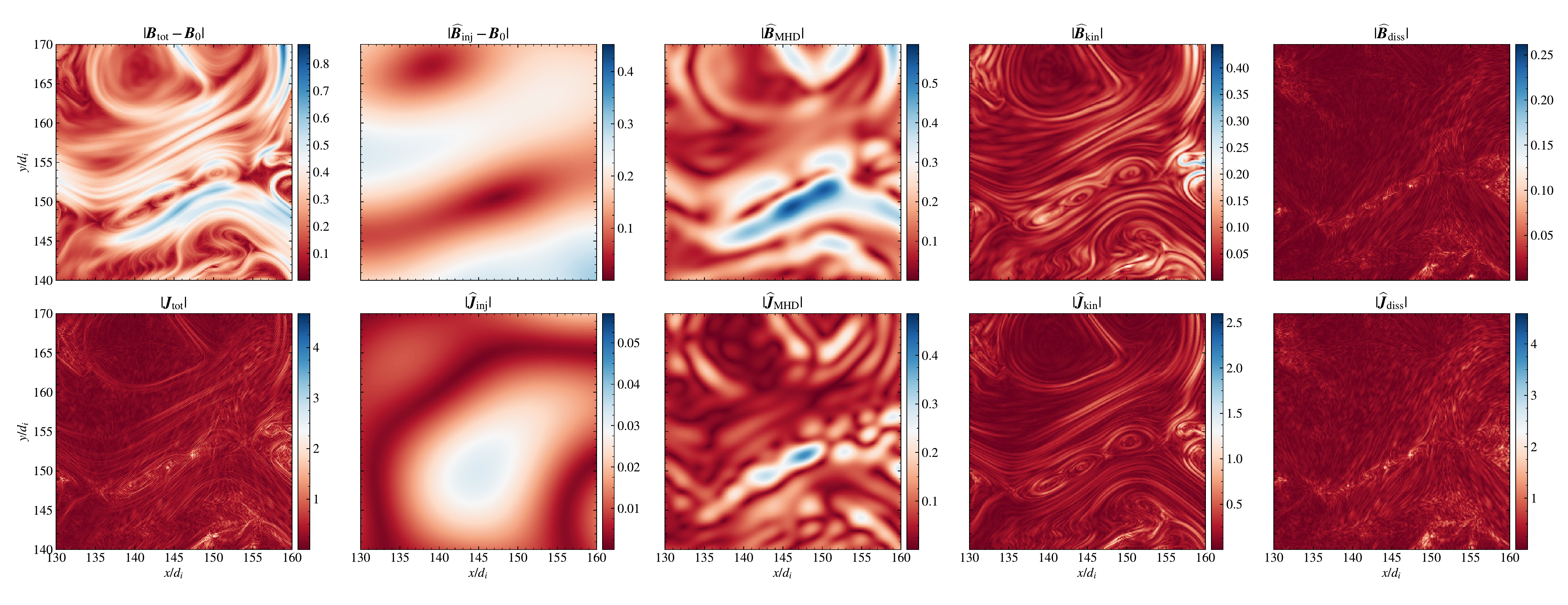}
  \caption{Top: amplitude of (from left to right) magnetic field fluctuations and corresponding aggregate IMFs of a subregion containing a current sheet undergoing plasmoid reconnection between two vortices, in the HPIC simulation and at the time of maximum turbulent activity. Bottom: same as the top panels but for the current density.}
\label{fig:aggregated_imfs_cs}
\end{figure}
As an example, Figure \ref{fig:aggregated_imfs_cs} displays the subregion $(x,y) \in [130,160]d_i \times [140,170] d_i $ of the HPIC simulation box, which hosts a chain of three plasmoids originated from the disruption of a reconnected  current sheet. The amplitude of the original magnetic field fluctuations and of its current density, $\boldsymbol{J} = \nabla\times\boldsymbol{B}$, is shown on the leftmost top and bottom panels respectively. There, we can distinguish the small plasmoids in the magnetic amplitudes, while the corresponding signal in the current density is almost swamped by the particle-per-cell (PPC) noise and by dissipation. The amplitudes of the aggregated IMFs (top) and of their associated current densities (bottom) are shown in the other panels. Now the three plasmoids are clearly visible in  $\IMF{\boldsymbol{B}}_\mathrm{kin}$ and $\IMF{\boldsymbol{J}}_\mathrm{kin}$, and their large-scale signature also appears at MHD scales. 

As espected, the aggregated IMFs also reveal that magnetic dissipation is mostly concentrated in strong current structures (high values of $\IMF{\boldsymbol{J}}_\mathrm{diss}$ are found in areas of high $\IMF{\boldsymbol{J}}_\mathrm{kin}$). Moreover, the highest dissipation (bright spots of $\IMF{\boldsymbol{J}}_\mathrm{diss}$) takes place at the X-points between the plasmoids, a typical feature of magnetic reconnection events.
All these morphological and physical multiscale properties, tipically difficult to isolate, are nicely and directly disentangled using MIF decomposition and in a straightforward manner.

Following \citet{1995Frisch}, a quantitative measure of the intermittency of a signal $f(x,y)$ can be provided  by measuring, at different frequencies $\kpj$,  the kurtosis 
\begin{equation}
 K(f_j^{>}) = \frac{\langle (f_j^{>})^4 \rangle}
 {\langle (f_j^{>})^2 \rangle^2}
\label{eq:kurtosis}
\end{equation}
where  $f_j^{>}(x,y)$ is a high-pass filtered signal of $f(x,y)$, only containing frequencies  $k_\perp \geq \kpj$, and where $\langle \rangle$ denote a spatial average. The signal $f(x,y)$ is intermittent if its kurtosis grows without bounds with frequency.
Using MIF decomposition, we can write the filtered signal as 
\begin{equation}
  f_j^{>}(x,y) = \sum_{k_\perp^{(i)} \geq \kpj} \IMF{f}_i(x,y),
\label{eq:highpassf}
  \end{equation}
by summing over all the IMF with an average frequency $k_\perp^{(i)} \geq \kpj$. 
We point out that Iterative Filtering based methods proved to be well suited to reconstruct the kurtosis, and more in general the multiscale statistical properties of nonstationary signals (including intermittent ones), as recently shown in \citet{2020stallone}.

Another quantity that measure the departure from a gaussian behavior is the Kullback-Leibler (KL) divergence \citep[e.g.,][]{2018GraneroBelinchon}. 
For a sample $X$ with probability density function (PDF) $p(x)$ of variance $\sigma_X^2$, the KL divergence is defined as
\begin{equation}
 \mathcal{K}_L (X) = H_G(X) - H(X),
\end{equation}
where 
\begin{equation}
 H (X)= -\int_\mathbb{R} p (x) \log p(x) \mathrm{d}x
\end{equation}
is the Shannon entropy of the sample, and $H_G = 0.5\log(2\pi e \sigma_X^2)$ is the Shannon entropy that $X$ would have if $p(x)$ were a Gaussian. $\mathcal{K}_L (X)$ is always positive, being zero if the PDF of the sample is a Gaussian distribution. Finally, the KL divergence $\mathcal{K}_L (f_j^{>})$ of $f(x,y)$ at a given frequency $\kpj$ is obtained by calculating the PDF and the variance of the values of $f_j^{>}(x,y)$.

\begin{figure}
  \includegraphics[width=.5\textwidth]{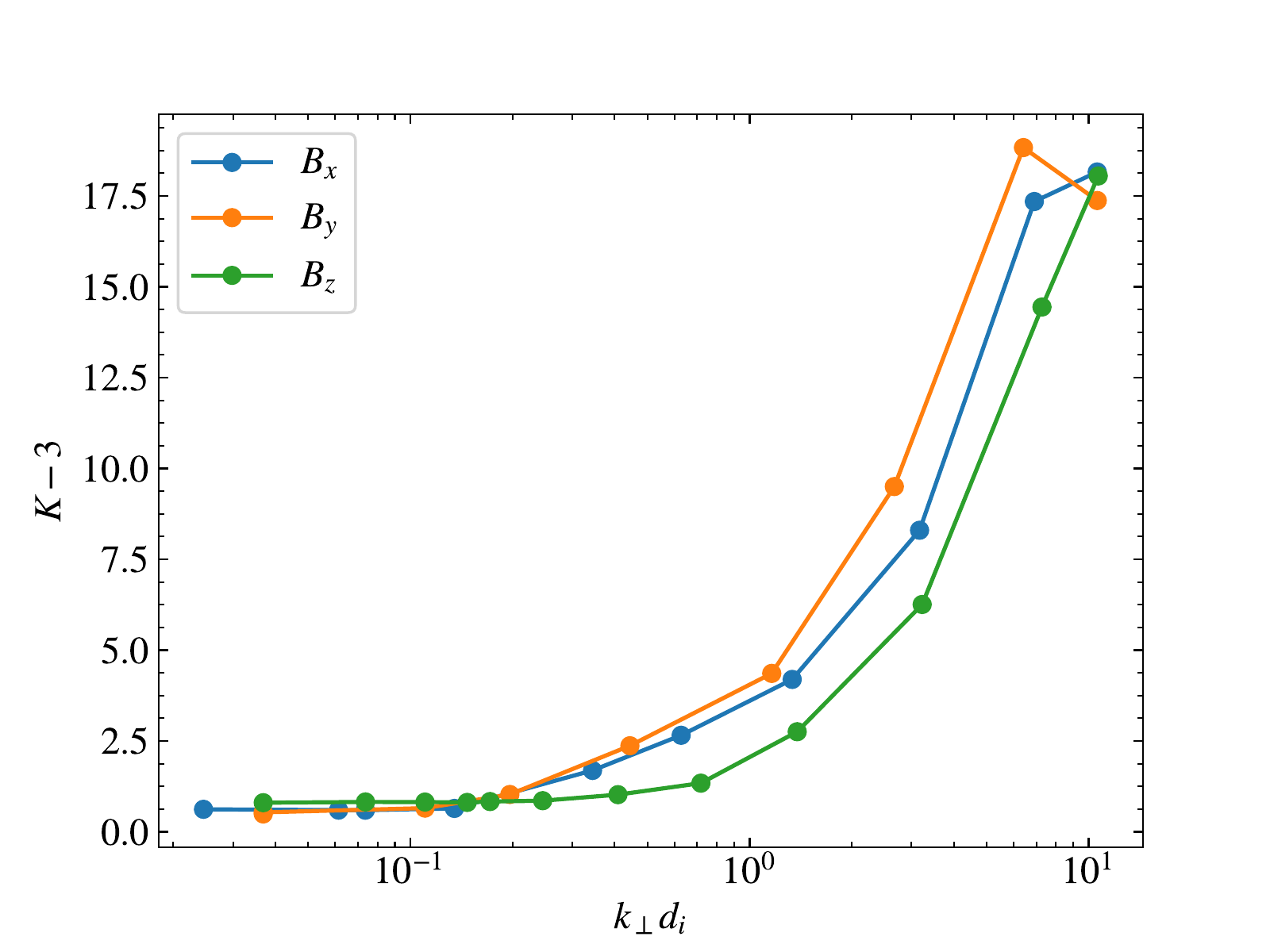}
  \includegraphics[width=.5\textwidth]{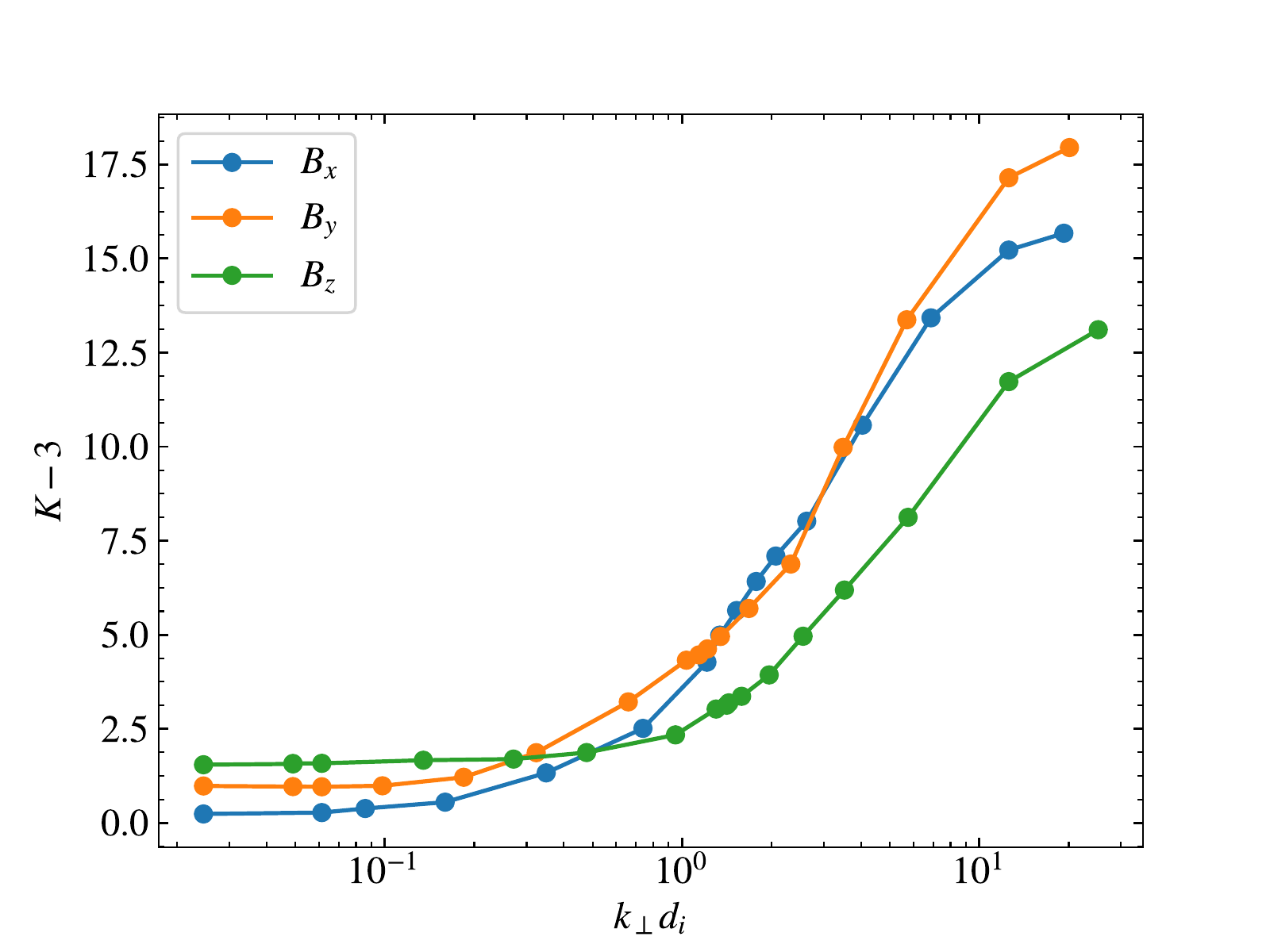}\\
   \includegraphics[width=.5\textwidth]{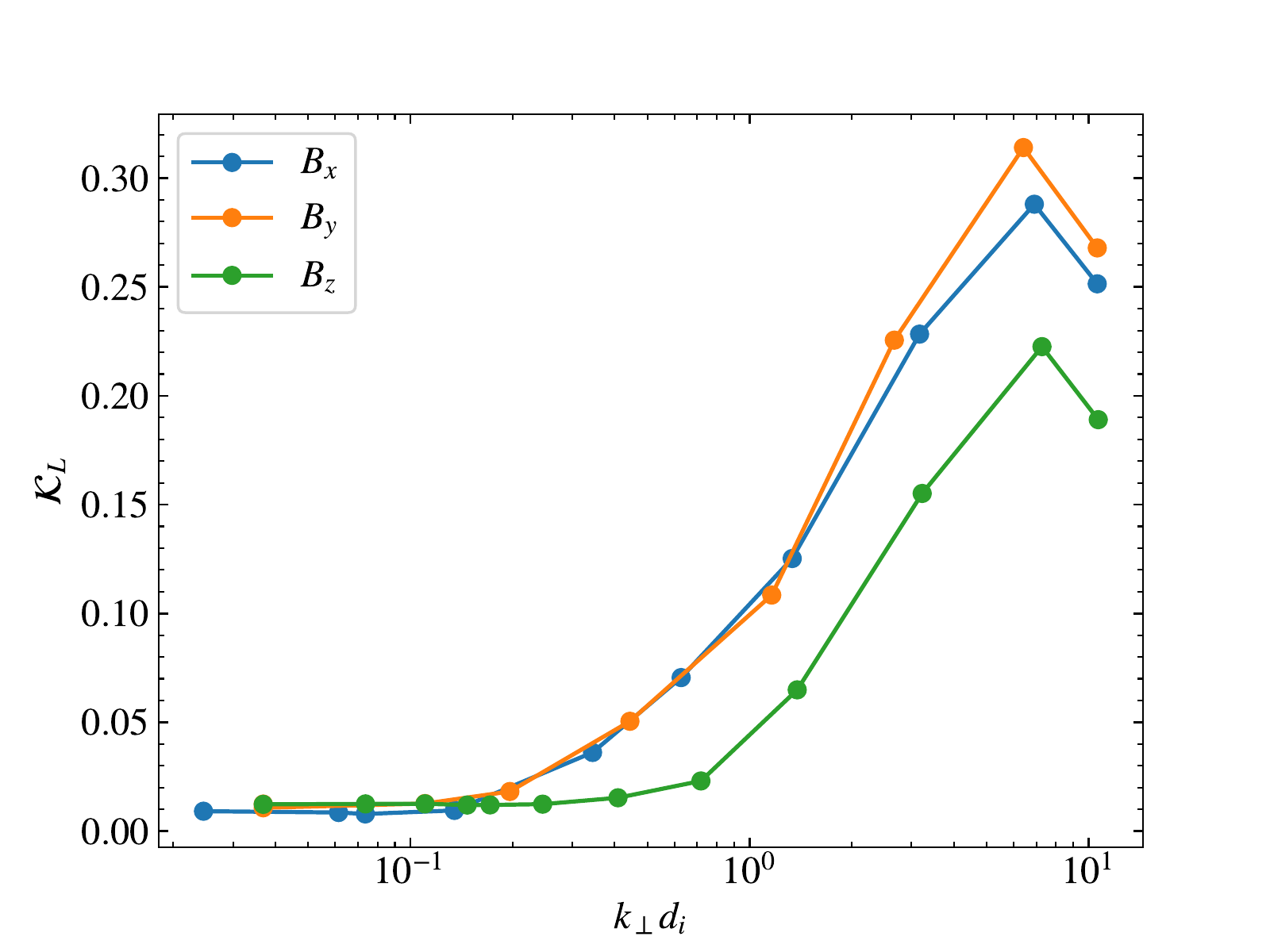}
   \includegraphics[width=.5\textwidth]{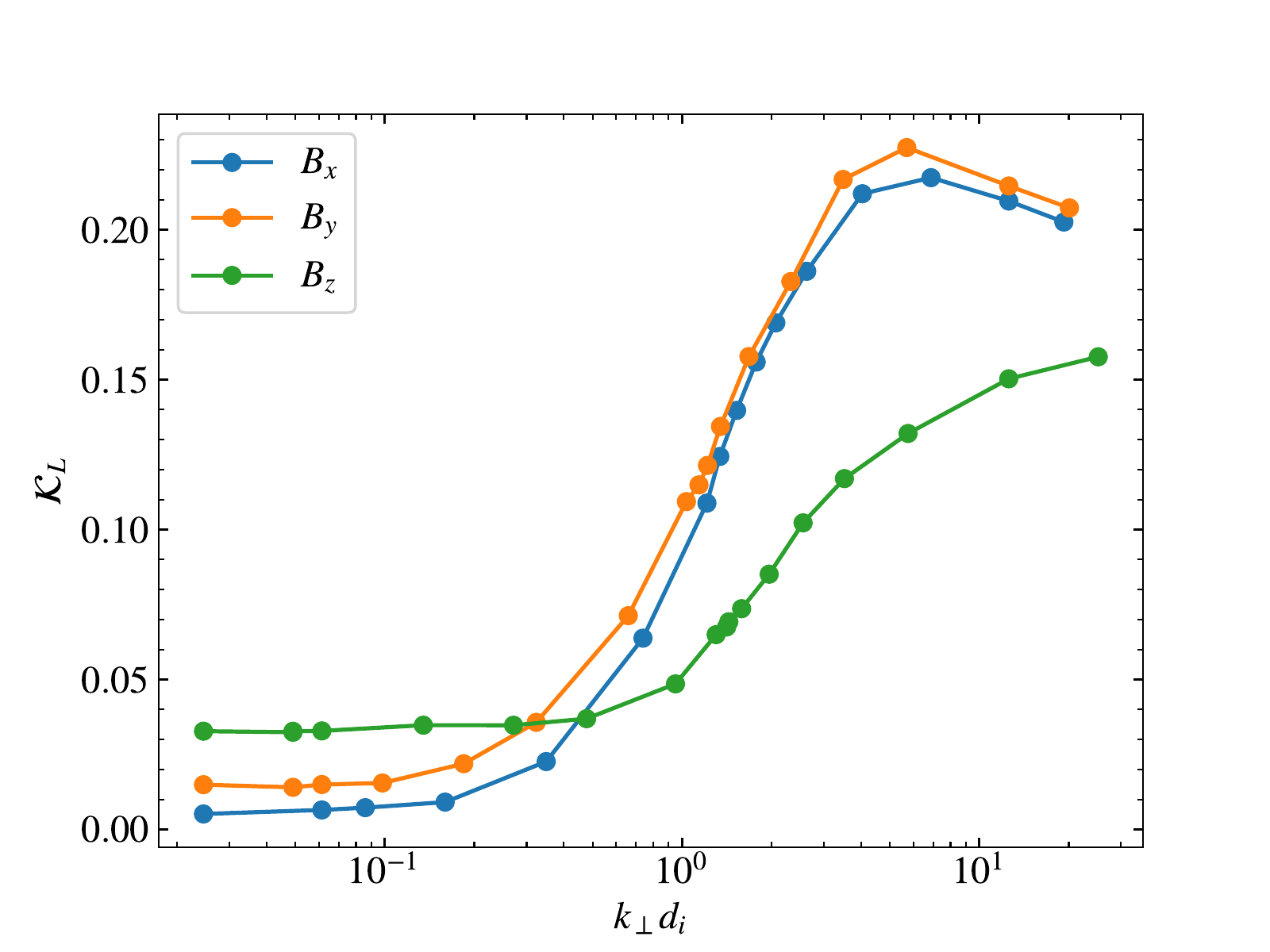}
  \caption{Excess kurtosis (top) and Kullback-Leibler divergence (bottom) of the magnetic field fluctuations as calculated from Eq. (\ref{eq:kurtosis}), by using the IMFs of the HMHD run (left) and of the HPIC run (right).}
\label{fig:kldiv}
\end{figure}

Figure \ref{fig:kldiv} shows the excess kurtosis, $K(f_j^{>}) - 3$, together with the KL divergence of the magnetic field components of the HMHD (left) and the HPIC (right) datasets. The results, in qualitative agreement with our previous findings \citep{2019papini_turb}, show a flat kurtosis at scales above the injection scale ($k_\perp d_i \lesssim 0.28$), denoting a self-similar behavior at those scales. The kurtosis of the perpendicular fluctuations start increasing in the MHD inertial range, then it steepens abruptly at kinetic scales, where Hall-current effects become important. This is in agreement with what observed in the aggregated IMFs: the presence of filamented networked structures at kinetic scales implies strong intermittency. 
The kurtosis of the parallel fluctuations $B_z$, instead,  remains constant in the inertial MHD range down to $k_\perp d_i = 1$, then abrutly increases with a behavior seimilar to the perpendicular fluctuations. We interpret such difference between parallel and perpendicular fluctuations at MHD scales as due to the particular setup we choose.
Alternatively, this could be the signature of the different behaviour of the two regimes: at the MHD scales, turbulence is leaded by Alfv\'enic-like 
fluctuations, which are predominately polarized in the direction perpendicular to $B_0$, while in the kinetic regime, dispersive effects couples the fluctuations with $B_z$.

 The simulations are inizialized with Alfv\'enic fluctuations in the perpendicular plane. In the inertial range, this causes the formation of a turbulent cascade in the perpendicular magnetic fluctuations, which then transition to kinetic scales. Instead, the magnetic power spectrum of parallel fluctuations \citep[see Fig. 5 of][]{2019papini_turb} show a cascade only at kinetic scales. Consequently, no intermittency develops until the disruption scales of current sheet (at around the ion inertial length $d_i$) are reached.
 
 The KL divergence (bottom panels of Figure \ref{fig:kldiv}) shows the same results. A departure from gaussian behavior ($\mathcal{K}_L >0$) is observed in the perpendicular fluctuations right below the injection scales and in the parallel fluctuations at kinetic scales. Interestingly, unlike the HMHD dataset, both the kurtosis and the KL divergence of $B_z$ in the HPIC dataset, although constant (i.e. not intermittent) at fluid scales ($k_\perp d_i<1$), are not zero, which denote a non-gaussian nature of the fluctuations. 
 We finally note that at the smallest scales, where dissipation kicks in, the KL divergence decreases in both the datasets, as expected.

\section{Discussion}

In this work, we investigated the magnetic multiscale properties of both fluid Hall-MHD and hybrid ion-kinetic electron-fluid simulations of plasma turbulence by means of Multidimensional Iterative Filtering (MIF), a novel technique for the decomposition of nonstationary multidimensional signals.
By exploiting our large-scale high-resolution numerical datasets,
we succesfully separated the four ranges of scales relevant to turbulence, namely the large injection scales, the inertial-range MHD scales, the sub-ion kinetic scales, and the dissipation scales. 
Moreover, we were able to reproduce the spectral and statistical properties of such regimes, while preserving the spatial information about morphology and localization of features and coherent structures, such as current sheets and vortices.

\subsection{Intermittency}

Our results confirm that plasma turbulence is an intrinsic multiscale phenomenon. 
In the inertial range, the energy cascade consists of more or less homegenously distributed and weakly intermittent magnetic fluctuations  (a slowly increasing kurtosis in the perpendicular fluctuations is observed). Ion kinetic scales are characterized by strongly-localized coherent structures organized in a filamented network, which show a high degree of intermittency.
Finally, when reaching the dissipation scales, the kurtosis tends to flatten and the KL-divergence decreases, suggesting that intermittency is switching off.
These results have been obtained by measuring the scale-dependent kurtosis and the KL-divergence of the magnetic field fluctuations, by means of a statistical analysis that exploits the MIF decomposition to calculate the high-pass filtered field $f_j^>$ contaning all spatial frequencies $k_\perp \geq \kpj$ (see Eq. \ref{eq:highpassf}).

It is instructive to compare these results with those obtained by using (i) Fourier transform in place of MIF decomposition to compute $f_j^>$ \citep[that is the exact definition of ][]{1995Frisch} and (ii) the magnetic field increments instead of $f_j^>$. 
The latter method has been applied  by \citet{2019papini_turb} to the same simulation dataset used in this work, and it is extensively employed both in solar wind and magnetosheath observations \citep[see, e.g.,][]{2007koga,2009chian,2013wu,2018bandyopadhyay}, as well as in numerical simulations \citep[e.g., ][]{2012wan,2015franci_b,2017haggerty}. 
\begin{figure}
\begin{center}
  \includegraphics[width=.7\textwidth]{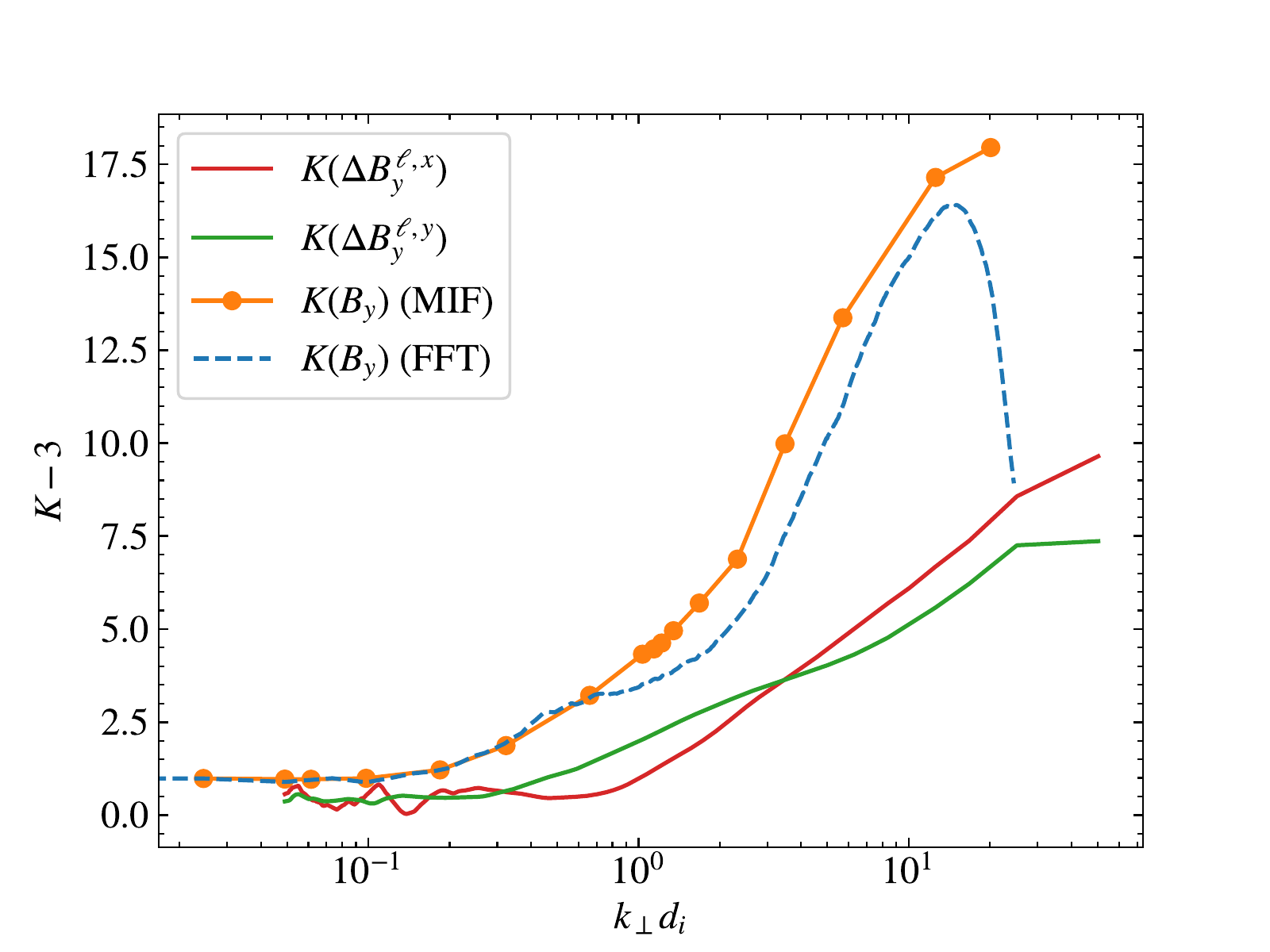}
  \end{center}
  \caption{Excess kurtosis of the $y$-component of the magnetic field fluctuations from the HPIC dataset, as calculated from the increments  along the $x$-direction ($\Delta B_y^{\ell,x}$) and along the $y$-direction ($\Delta B_y^{\ell,y}$). The excess kurtosis as given by Eq. \ref{eq:kurtosis}, by using MIF (solid orange curve and circles) and FFT (blue dashed curve) methods to calculate the high-pass filtered function $f_j^>$ of $B_y$, is also shown.}
\label{fig:kurt_dby}
\end{figure}

In Figure \ref{fig:kurt_dby} we report, for the HPIC run, the kurtosis of the increments $\Delta B_y^{\ell,y} = B_y(x,y+\ell) - B_y(x,y)$ (green curve) and $\Delta B_y^{\ell,x}=B_y(x+\ell,y)-B_y(x,y)$ (red curve), respectively parallel and perpendicular (in the plane) to the direction of the $B_y$ component (where $\ell = 2\pi/k_\perp$).
We also report the kurtosis obtained from the MIF decomposition of $B_y$ (already shown in the top-right panel of Figure \ref{fig:kldiv}) and the one obtained using Fourier transform to calculate the high-pass filtered field $f_j^>$ from $B_y$.
Overall, the results are in qualitative agreement. 
There are, however, some noticeable differences. The kurtosis of the increments $\Delta B_y^{\ell,x}$ is almost constant in the inertial range, then linearly increases starting from $k_\perp d_i \simeq 1$, i.e., at the scales where Hall currents become important.
Instead, $K(\Delta B_y^{\ell,y})$ linearly increases already at the injection scales, but with a smaller slope than $K(\Delta B_y^{\ell,x})$.
The kurtosis calculated with MIF, although it is roughly twice as larger, reproduces the properties of both $K(\Delta B_y^{\ell,y})$ and $K(\Delta B_y^{\ell,x})$, following the former at the large scales down to the spectral break $k_\perp d_i \simeq 2$ and then steepening at the scales where $K(\Delta B_y^{\ell,x}) > K(\Delta B_y^{\ell,y})$.

Such differences may be explained by considering the following argument. 
Firstly, the increments $\Delta B_y^{\ell,y}\propto\partial_y B_y$ and $\Delta B_y^{\ell,x}\propto\partial_x B_y$ are proportional to the components of the magnetic field gradient parallel and perpendicular to the magnetic field direction, respectively. 
Secondly, localized and elongated magnetic structures, such as the thin current sheets between vortices that we observed, are aligned with the magnetic field. Therefore, $K(\Delta B_y^{\ell,x})$ is sensitive to the thickness of such structures (which is of the order of $d_i$) while $K(\Delta B_y^{\ell,y})$ probes their length (that is comparable to the size of the largest vortices). This may explain the observed increase in the parallel and perpendicular kurtosis at around those scales. 
We finally remark that the kurtosis calculated using the Fourier and the MIF decomposition are in remarkable agreement, although the former 
shows smaller values at high frequencies and even bring to a sudden decrease at the highest frequencies
(because of the increasing inability of the Fourier high-pass filtered field to localize structures, as $k_\perp$ increases). Analogous results are obtained for the $x$-component of the magnetic field.

Overall, the multiscale statistical properties we recovered are consistent with the findings of \citet{2019alberti} (hereafter AL19). They measured the multifractal nature of solar wind turbulence by using CLUSTER data and found increasing levels of intermittency in the MHD/inertial range, with a tendency toward a non-intermittent/monofractal behavior at dissipation scales. 
There is, however, an important difference. In the ion kinetic range, where AL19 observe a monofractal behavior, we find high levels of intermittency, which denote  a multifractal nature of the fluctuations. We are not certain whether such difference is due to the particular dataset chosen by AL19 (a fast solar wind stream) or by our HMHD and HPIC models. 
We note, however, that at ion kinetic scales AL19 measure magnetic power spectra with a slope of $\sim -5/2$, different from the one we report in our simulations ($-3$) and also from other solar wind conditions.
Further investigation pursuing this path is currently underway, in order to assess the multifractal properties of our numerical simulations.

\subsection{Reconnection and enhanced dissipation}

Our multiscale analysis confirms that magnetic field dissipation is mostly concentrated in the filamented magnetic network, especially at the X-point of reconnecting current sheets.
This is in agreement with previous studies of plasma turbulence that measured the scale-to-scale energy transfer by means of scale-filtering approaches \citep{2017yang,2018camporeale, 2019kuzzay}.
The filamented magnetic network observed at kinetic scales is somewhat reminiscent of the vortex filaments in hydrodynamic turbulence \citep[e.g., ][]{1992kida,1994moffatt}, and consistent with the fact that areas of both high magnetic gradients and vorticities are correlated \citep{2016franci_proc,2019kuzzay}, especially at the reconnection sites, which produce high levels of vorticity \citep[e.g.,][]{2016widmer}.
Furthermore, the physical and geometrical features typical of magnetic reconnection were easily disentangled and identified, even when the signal of the structure was swamped by PPC noise and dissipation (see Figure \ref{fig:aggregated_imfs_cs}). In this context, Multidimensional Iterative Filtering also stands as a powerful tool for automatically identifying and removing noise from the physical signal of interest.

Currently, we are conducting a time-frequency analysis of our simulations, by employing IF. The aim is to confirm whether the turbulent dynamics at kinetic scales is wavelike (e.g. mediated by kinetic alfv\'en wavelike interactions) or it is due to the presence of structures, as the results of this work seems to suggest.
Overall, Multidimensional Iterative Filtering is a promising technique to support the study of plasma turbulence and its properties, with many potential applications in both numerical simulations and  spacecraft observations.

\begin{acknowledgments}
E. Papini and S. Landi thank T. Alberti and G. Consolini for useful discussion.
M. Piersanti thanks the Italian Space Agency for the financial support under the contract ASI ”LIMADOU scienza” n$^{\circ}$ 2016-16-H0."
This research was partially supported by the UK Science and Technology Facilities Council (STFC) grant ST/P000622/.
This work was supported by the Programme National PNST of CNRS/INSU co-funded by CNES
P. Hellinger acknowledges grant 18-08861S of the Czech Science Foundation.
We acknowledge partial funding by ``Fondazione Cassa di Risparmio di Firenze'' under the project HIPERCRHEL.
The authors acknowledge the ``Accordo Quadro INAF-CINECA (2017)'', for the availability of high performance computing resources and support, PRACE for awarding access to the resource Cartesius based in the Netherlands at SURFsara through the DECI-13 (Distributed European Computing Initiative) call (project HybTurb3D), and CINECA for awarding access to HPC resources under the ISCRA initiative (grants HP10B2DRR4 and HP10C2EARF). 
\end{acknowledgments}

\bibliographystyle{jpp}

\bibliography{jppbiblio}

\begin{thebibliography}{72}
\expandafter\ifx\csname natexlab\endcsname\relax\def\natexlab#1{#1}\fi
\def\au#1{#1} \def\ed#1{#1} \def\yr#1{#1}\def\at#1{#1}\def\jt#1{\textit{#1}}
  \def\bt#1{#1}\def\bvol#1{\textbf{#1}} \def\vol#1{#1} \def\pg#1{#1}
  \def\publ#1{#1}\def\arxiv#1{#1}\def\org#1{#1}\def\st#1{\textit{#1}}

\bibitem[{Alberti} {\em et~al.\/}(2019){Alberti}, {Consolini}, {Carbone},
  {Yordanova}, {Marcucci} \& {De Michelis}]{2019alberti}
{\sc \au{{Alberti}, T.}, \au{{Consolini}, G.}, \au{{Carbone}, V.},
  \au{{Yordanova}, E.}, \au{{Marcucci}, M.} \& \au{{De Michelis}, P.}}
  \yr{2019}  \at{{Multifractal and Chaotic Properties of Solar Wind at MHD and
  Kinetic Domains: An Empirical Mode Decomposition Approach}}.  \jt{Entropy}
  \bvol{21}~(3),  \pg{320}.

\bibitem[{Bandyopadhyay} {\em et~al.\/}(2018){Bandyopadhyay}, {Chasapis},
  {Chhiber}, {Parashar}, {Maruca}, {Matthaeus}, {Schwartz}, {Eriksson}, {Le
  Contel}, {Breuillard}, {Burch}, {Moore}, {Pollock}, {Giles}, {Paterson},
  {Dorelli}, {Gershman}, {Torbert}, {Russell} \&
  {Strangeway}]{2018bandyopadhyay}
{\sc \au{{Bandyopadhyay}, R.}, \au{{Chasapis}, A.}, \au{{Chhiber}, R.},
  \au{{Parashar}, T.~N.}, \au{{Maruca}, B.~A.}, \au{{Matthaeus}, W.~H.},
  \au{{Schwartz}, S.~J.}, \au{{Eriksson}, S.}, \au{{Le Contel}, O.},
  \au{{Breuillard}, H.}, \au{{Burch}, J.~L.}, \au{{Moore}, T.~E.},
  \au{{Pollock}, C.~J.}, \au{{Giles}, B.~L.}, \au{{Paterson}, W.~R.},
  \au{{Dorelli}, J.}, \au{{Gershman}, D.~J.}, \au{{Torbert}, R.~B.},
  \au{{Russell}, C.~T.} \& \au{{Strangeway}, R.~J.}} \yr{2018}  \at{{Solar Wind
  Turbulence Studies Using MMS Fast Plasma Investigation Data}}.
  \jt{Astrophys. J.}  \bvol{866}~(2),  \pg{81}.

\bibitem[{Bertello} {\em et~al.\/}(2018){Bertello}, {Piersanti}, {Candidi},
  {Diego} \& {Ubertini}]{2018bertello}
{\sc \au{{Bertello}, I.}, \au{{Piersanti}, M.}, \au{{Candidi}, M.},
  \au{{Diego}, P.} \& \au{{Ubertini}, P.}} \yr{2018}  \at{{Electromagnetic
  field observations by the DEMETER satellite in connection with the 2009
  L'Aquila earthquake}}.  \jt{Annales Geophysicae}  \bvol{36}~(5),
  \pg{1483--1493}.

\bibitem[{Boldyrev} \& {Perez}(2012)]{2012boldyrev}
{\sc \au{{Boldyrev}, S.} \& \au{{Perez}, J.~C.}} \yr{2012}  \at{{Spectrum of
  Kinetic-Alfv{\'e}n Turbulence}}.  \jt{Astrophys. J. Lett.}  \bvol{758},
  \pg{L44}.

\bibitem[{Bruno} {\em et~al.\/}(2001){Bruno}, {Carbone}, {Veltri},
  {Pietropaolo} \& {Bavassano}]{2001bruno}
{\sc \au{{Bruno}, R.}, \au{{Carbone}, V.}, \au{{Veltri}, P.},
  \au{{Pietropaolo}, E.} \& \au{{Bavassano}, B.}} \yr{2001}  \at{{Identifying
  intermittency events in the solar wind}}.  \jt{Planetary and Space Science}
  \bvol{49}~(12),  \pg{1201--1210}.

\bibitem[{Bruno} {\em et~al.\/}(2009){Bruno}, {Carbone}, {V{\"o}r{\"o}s},
  {D'Amicis}, {Bavassano}, {Cattaneo}, {Mura}, {Milillo}, {Orsini}, {Veltri},
  {Sorriso-Valvo}, {Zhang}, {Biernat}, {Rucker}, {Baumjohann},
  {Jankovi{\v{c}}ov{\'a}} \& {Kov{\'a}cs}]{2009bruno}
{\sc \au{{Bruno}, R.}, \au{{Carbone}, V.}, \au{{V{\"o}r{\"o}s}, Z.},
  \au{{D'Amicis}, R.}, \au{{Bavassano}, B.}, \au{{Cattaneo}, M.~B.},
  \au{{Mura}, A.}, \au{{Milillo}, A.}, \au{{Orsini}, S.}, \au{{Veltri}, P.},
  \au{{Sorriso-Valvo}, L.}, \au{{Zhang}, T.}, \au{{Biernat}, H.}, \au{{Rucker},
  H.}, \au{{Baumjohann}, W.}, \au{{Jankovi{\v{c}}ov{\'a}}, D.} \&
  \au{{Kov{\'a}cs}, P.}} \yr{2009}  \at{{Coordinated Study on Solar Wind
  Turbulence During the Venus-Express, ACE and Ulysses Alignment of August
  2007}}.  \jt{Earth Moon and Planets}  \bvol{104}~(1-4),  \pg{101--104}.

\bibitem[{Camporeale} {\em et~al.\/}(2018){Camporeale}, {Sorriso-Valvo},
  {Califano} \& {Retin{\`o}}]{2018camporeale}
{\sc \au{{Camporeale}, E.}, \au{{Sorriso-Valvo}, L.}, \au{{Califano}, F.} \&
  \au{{Retin{\`o}}, A.}} \yr{2018}  \at{{Coherent Structures and Spectral
  Energy Transfer in Turbulent Plasma: A Space-Filter Approach}}.  \jt{Phys.
  Rev. Lett.}  \bvol{120}~(12),  \pg{125101}.

\bibitem[{Cerri} \& {Califano}(2017)]{2017cerri}
{\sc \au{{Cerri}, S.~S.} \& \au{{Califano}, F.}} \yr{2017}  \at{{Reconnection
  and small-scale fields in 2D-3V hybrid-kinetic driven turbulence
  simulations}}.  \jt{New Journal of Physics}  \bvol{19}~(2),  \pg{025007}.

\bibitem[{Chang} {\em et~al.\/}(2004){Chang}, {Tam} \& {Wu}]{2004chang}
{\sc \au{{Chang}, T.}, \au{{Tam}, S. W.~Y.} \& \au{{Wu}, C.-C.}} \yr{2004}
  \at{{Complexity induced anisotropic bimodal intermittent turbulence in space
  plasmas}}.  \jt{Physics of Plasmas}  \bvol{11}~(4),  \pg{1287--1299}.

\bibitem[{Chen} {\em et~al.\/}(2013){Chen}, {Bale}, {Salem} \&
  {Maruca}]{2013chen}
{\sc \au{{Chen}, C.~H.~K.}, \au{{Bale}, S.~D.}, \au{{Salem}, C.~S.} \&
  \au{{Maruca}, B.~A.}} \yr{2013}  \at{{Residual Energy Spectrum of Solar Wind
  Turbulence}}.  \jt{Astrophys. J.}  \bvol{770},  \pg{125}.

\bibitem[{Chen} \& {Boldyrev}(2017)]{2017chen}
{\sc \au{{Chen}, C.~H.~K.} \& \au{{Boldyrev}, S.}} \yr{2017}  \at{{Nature of
  Kinetic Scale Turbulence in the Earth's Magnetosheath}}.  \jt{Astrophys. J.}
  \bvol{842},  \pg{122}.

\bibitem[{Chian} \& {Miranda}(2009)]{2009chian}
{\sc \au{{Chian}, A.~C.~L.} \& \au{{Miranda}, R.~A.}} \yr{2009}  \at{{Cluster
  and ACE observations of phase synchronization in intermittent magnetic field
  turbulence: a comparative study of shocked and unshocked solar wind}}.
  \jt{Annales Geophysicae}  \bvol{27}~(4),  \pg{1789--1801}.

\bibitem[Cicone(2020)]{2020cicone}
{\sc \au{Cicone, A.}} \yr{2020}  \at{Iterative filtering as a direct method for
  the decomposition of nonstationary signals}.  \jt{Numerical Algorithms}
  \pg{pp. 1--17}.

\bibitem[Cicone \& Dell’Acqua(2020)]{2020cicone_boundary}
{\sc \au{Cicone, A.} \& \au{Dell’Acqua, P.}} \yr{2020}  \at{Study of boundary
  conditions in the iterative filtering method for the decomposition of
  nonstationary signals}.  \jt{Journal of Computational and Applied
  Mathematics}  \bvol{373},  \pg{112248}.

\bibitem[Cicone {\em et~al.\/}(2016)Cicone, Liu \& Zhou]{2016cicone}
{\sc \au{Cicone, A.}, \au{Liu, J.} \& \au{Zhou, H.}} \yr{2016}  \at{Adaptive
  local iterative filtering for signal decomposition and instantaneous
  frequency analysis}.  \jt{Applied and Computational Harmonic Analysis}
  \bvol{41}~(2),  \pg{384 -- 411}, sparse Representations with Applications in
  Imaging Science, Data Analysis, and Beyond, Part II.

\bibitem[Cicone \& Zhou(2017)]{2017Cicone}
{\sc \au{Cicone, A.} \& \au{Zhou, H.}} \yr{2017}  \at{Multidimensional
  iterative filtering method for the decomposition of high–dimensional
  non–stationary signals}.  \jt{Numerical Mathematics: Theory, Methods and
  Applications}  \bvol{10}~(2),  \pg{278–298}.

\bibitem[{Cicone} \& {Zhou}(2020)]{2020cicone_fft}
{\sc \au{{Cicone}, A.} \& \au{{Zhou}, H.}} \yr{2020}  \at{{Numerical Analysis
  for Iterative Filtering with New Efficient Implementations Based on FFT}}.
  \jt{submitted} ,  \arxiv{arXiv: 1802.01359}.

\bibitem[Consolini {\em et~al.\/}(2005)Consolini, Chang \& Lui]{2005consolini}
{\sc \au{Consolini, G.}, \au{Chang, T.} \& \au{Lui, A. T.~Y.}} \yr{2005} {\em
  Complexity and Topological Disorder in the Earth's Magnetotail Dynamics\/},
  \pg{pp. 51--69}.  \publ{Dordrecht: Springer Netherlands}.

\bibitem[{Del Sarto} \& {Pegoraro}(2018)]{DelSarto_Pegoraro_2018}
{\sc \au{{Del Sarto}, D.} \& \au{{Pegoraro}, F.}} \yr{2018}  \at{{Shear-induced
  pressure anisotropization and correlation with fluid vorticity in a low
  collisionality plasma}}.  \jt{Mon. Not. R. Astron.}  \bvol{475},
  \pg{181--192}.

\bibitem[{Franci} {\em et~al.\/}(2017){Franci}, {Cerri}, {Califano}, {Landi},
  {Papini}, {Verdini}, {Matteini}, {Jenko} \& {Hellinger}]{2017franci}
{\sc \au{{Franci}, L.}, \au{{Cerri}, S.~S.}, \au{{Califano}, F.}, \au{{Landi},
  S.}, \au{{Papini}, E.}, \au{{Verdini}, A.}, \au{{Matteini}, L.}, \au{{Jenko},
  F.} \& \au{{Hellinger}, P.}} \yr{2017}  \at{{Magnetic Reconnection as a
  Driver for a Sub-ion-scale Cascade in Plasma Turbulence}}.  \jt{Astrophys. J.
  Lett.}  \bvol{850},  \pg{L16}.

\bibitem[Franci {\em et~al.\/}(2018{\natexlab{{\em a\/}}})Franci, Hellinger,
  Guarrasi, Chen, Papini, Verdini, Matteini \& Landi]{2018franci}
{\sc \au{Franci, L.}, \au{Hellinger, P.}, \au{Guarrasi, M.}, \au{Chen, C.
  H.~K.}, \au{Papini, E.}, \au{Verdini, A.}, \au{Matteini, L.} \& \au{Landi,
  S.}} \yr{2018{\natexlab{{\em a\/}}}}  \at{Three-dimensional simulations of
  solar wind turbulence with the hybrid code camelia}.  \jt{J. Phys.: Conf.
  Series}  \bvol{1031}~(1),  \pg{012002}.

\bibitem[{Franci} {\em et~al.\/}(2016{\natexlab{{\em a\/}}}){Franci},
  {Hellinger}, {Matteini}, {Verdini} \& {Landi}]{2016bFranci}
{\sc \au{{Franci}, L.}, \au{{Hellinger}, P.}, \au{{Matteini}, L.},
  \au{{Verdini}, A.} \& \au{{Landi}, S.}} \yr{2016{\natexlab{{\em a\/}}}}
  {Two-dimensional hybrid simulations of kinetic plasma turbulence: Current and
  vorticity vs proton temperature}.  \bt{In {\em AIP Conf. Series\/}},  \st{AIP
  Conf. Series},  \vol{vol. 1720},  \pg{p. 040003},  \arxiv{arXiv: 1604.03040}.

\bibitem[{Franci} {\em et~al.\/}(2016{\natexlab{{\em b\/}}}){Franci},
  {Hellinger}, {Matteini}, {Verdini} \& {Landi}]{2016franci_proc}
{\sc \au{{Franci}, L.}, \au{{Hellinger}, P.}, \au{{Matteini}, L.},
  \au{{Verdini}, A.} \& \au{{Landi}, S.}} \yr{2016{\natexlab{{\em b\/}}}}
  {Two-dimensional hybrid simulations of kinetic plasma turbulence: Current and
  vorticity vs proton temperature}.  \bt{In {\em American Institute of Physics
  Conference Series\/}},  \st{American Institute of Physics Conference Series},
   \vol{vol. 1720},  \pg{p. 040003}.

\bibitem[{Franci} {\em et~al.\/}(2015{\natexlab{{\em a\/}}}){Franci}, {Landi},
  {Matteini}, {Verdini} \& {Hellinger}]{2015franci_b}
{\sc \au{{Franci}, L.}, \au{{Landi}, S.}, \au{{Matteini}, L.}, \au{{Verdini},
  A.} \& \au{{Hellinger}, P.}} \yr{2015{\natexlab{{\em a\/}}}}
  \at{{High-resolution Hybrid Simulations of Kinetic Plasma Turbulence at
  Proton Scales}}.  \jt{Astrophys. J.}  \bvol{812},  \pg{21}.

\bibitem[{Franci} {\em et~al.\/}(2016{\natexlab{{\em c\/}}}){Franci}, {Landi},
  {Matteini}, {Verdini} \& {Hellinger}]{2016franci}
{\sc \au{{Franci}, L.}, \au{{Landi}, S.}, \au{{Matteini}, L.}, \au{{Verdini},
  A.} \& \au{{Hellinger}, P.}} \yr{2016{\natexlab{{\em c\/}}}}  \at{{Plasma
  Beta Dependence of the Ion-scale Spectral Break of Solar Wind Turbulence:
  High-resolution 2D Hybrid Simulations}}.  \jt{Astrophys. J.}  \bvol{833},
  \pg{91}.

\bibitem[Franci {\em et~al.\/}(2018{\natexlab{{\em b\/}}})Franci, Landi,
  Verdini, Matteini \& Hellinger]{Franci_al_2018}
{\sc \au{Franci, L.}, \au{Landi, S.}, \au{Verdini, A.}, \au{Matteini, L.} \&
  \au{Hellinger, P.}} \yr{2018{\natexlab{{\em b\/}}}}  \at{{Solar Wind
  Turbulent Cascade from MHD to Sub-ion Scales: Large-size 3D Hybrid
  Particle-in-cell Simulations}}.  \jt{Astrophys. J.}  \bvol{853},  \pg{26}.

\bibitem[{Franci} {\em et~al.\/}(2019{\natexlab{{\em a\/}}}){Franci},
  {Stawarz}, {Papini}, {Hellinger}, {Nakamura}, {Burgess}, {Land i}, {Verdini},
  {Matteini}, {Ergun}, {Le Contel} \& {Lindqvist}]{2019franci}
{\sc \au{{Franci}, L.}, \au{{Stawarz}, J.~E.}, \au{{Papini}, E.},
  \au{{Hellinger}, P.}, \au{{Nakamura}, T.}, \au{{Burgess}, D.}, \au{{Land i},
  S.}, \au{{Verdini}, A.}, \au{{Matteini}, L.}, \au{{Ergun}, R.}, \au{{Le
  Contel}, O.} \& \au{{Lindqvist}, P.-A.}} \yr{2019{\natexlab{{\em a\/}}}}
  \at{{Modeling Kelvin-Helmholtz instability-driven turbulence with hybrid
  simulations of Alfv{\'e}nic turbulence}}.  \jt{arXiv e-prints}  \pg{p.
  arXiv:1911.07370},  \arxiv{arXiv: 1911.07370}.

\bibitem[{Franci} {\em et~al.\/}(2019{\natexlab{{\em b\/}}}){Franci},
  {Stawarz}, {Papini}, {Hellinger}, {Nakamura}, {Burgess}, {Land i}, {Verdini},
  {Matteini}, {Ergun}, {Le Contel} \& {Lindqvist}]{2020franci}
{\sc \au{{Franci}, L.}, \au{{Stawarz}, J.~E.}, \au{{Papini}, E.},
  \au{{Hellinger}, P.}, \au{{Nakamura}, T.}, \au{{Burgess}, D.}, \au{{Land i},
  S.}, \au{{Verdini}, A.}, \au{{Matteini}, L.}, \au{{Ergun}, R.}, \au{{Le
  Contel}, O.} \& \au{{Lindqvist}, P.-A.}} \yr{2019{\natexlab{{\em b\/}}}}
  \at{{Modeling Kelvin-Helmholtz instability-driven turbulence with hybrid
  simulations of Alfv{\'e}nic turbulence}}.  \jt{submitted}  \pg{p.
  arXiv:1911.07370},  \arxiv{arXiv: 1911.07370}.

\bibitem[{Franci} {\em et~al.\/}(2015{\natexlab{{\em b\/}}}){Franci},
  {Verdini}, {Matteini}, {Landi} \& {Hellinger}]{2015franci_a}
{\sc \au{{Franci}, L.}, \au{{Verdini}, A.}, \au{{Matteini}, L.}, \au{{Landi},
  S.} \& \au{{Hellinger}, P.}} \yr{2015{\natexlab{{\em b\/}}}}  \at{{Solar Wind
  Turbulence from MHD to Sub-ion Scales: High-resolution Hybrid Simulations}}.
  \jt{Astrophys. J. Lett.}  \bvol{804},  \pg{L39}.

\bibitem[{Frisch}(1995)]{1995Frisch}
{\sc \au{{Frisch}, U.}} \yr{1995} {\em {Turbulence. The legacy of A.N.
  Kolmogorov}\/}.  \publ{{Cambridge University Press}}.

\bibitem[{Granero-Belinch{\'o}n} {\em et~al.\/}(2018){Granero-Belinch{\'o}n},
  {Roux} \& {Garnier}]{2018GraneroBelinchon}
{\sc \au{{Granero-Belinch{\'o}n}, C.}, \au{{Roux}, S.~G.} \& \au{{Garnier},
  N.~B.}} \yr{2018}  \at{{Kullback-Leibler divergence measure of intermittency:
  Application to turbulence}}.  \jt{Phys. Rev. E}  \bvol{97}~(1),  \pg{013107}.

\bibitem[{Haggerty} {\em et~al.\/}(2017){Haggerty}, {Parashar}, {Matthaeus},
  {Shay}, {Yang}, {Wan}, {Wu} \& {Servidio}]{2017haggerty}
{\sc \au{{Haggerty}, C.~C.}, \au{{Parashar}, T.~N.}, \au{{Matthaeus}, W.~H.},
  \au{{Shay}, M.~A.}, \au{{Yang}, Y.}, \au{{Wan}, M.}, \au{{Wu}, P.} \&
  \au{{Servidio}, S.}} \yr{2017}  \at{{Exploring the statistics of magnetic
  reconnection X-points in kinetic particle-in-cell turbulence}}.  \jt{Physics
  of Plasmas}  \bvol{24}~(10),  \pg{102308}.

\bibitem[{Hellinger} {\em et~al.\/}(2018){Hellinger}, {Verdini}, {Landi},
  {Franci} \& {Matteini}]{2018hellinger}
{\sc \au{{Hellinger}, P.}, \au{{Verdini}, A.}, \au{{Landi}, S.}, \au{{Franci},
  L.} \& \au{{Matteini}, L.}} \yr{2018}  \at{{von K{\'a}rm{\'a}n-Howarth
  Equation for Hall Magnetohydrodynamics: Hybrid Simulations}}.  \jt{Astrophys.
  J. Lett.}  \bvol{857},  \pg{L19}.

\bibitem[{Horbury} {\em et~al.\/}(2008){Horbury}, {Forman} \&
  {Oughton}]{2008horbury}
{\sc \au{{Horbury}, T.~S.}, \au{{Forman}, M.} \& \au{{Oughton}, S.}} \yr{2008}
  \at{{Anisotropic Scaling of Magnetohydrodynamic Turbulence}}.  \jt{\prl}
  \bvol{101}~(17),  \pg{175005},  \arxiv{arXiv: 0807.3713}.

\bibitem[{Howes} {\em et~al.\/}(2008){Howes}, {Cowley}, {Dorland}, {Hammett},
  {Quataert} \& {Schekochihin}]{2008howes}
{\sc \au{{Howes}, G.~G.}, \au{{Cowley}, S.~C.}, \au{{Dorland}, W.},
  \au{{Hammett}, G.~W.}, \au{{Quataert}, E.} \& \au{{Schekochihin}, A.~A.}}
  \yr{2008}  \at{{A model of turbulence in magnetized plasmas: Implications for
  the dissipation range in the solar wind}}.  \jt{Journal of Geophysical
  Research (Space Physics)}  \bvol{113}~(A5),  \pg{A05103}.

\bibitem[{Howes} {\em et~al.\/}(2011){Howes}, {Tenbarge}, {Dorland},
  {Quataert}, {Schekochihin}, {Numata} \& {Tatsuno}]{2011howes}
{\sc \au{{Howes}, G.~G.}, \au{{Tenbarge}, J.~M.}, \au{{Dorland}, W.},
  \au{{Quataert}, E.}, \au{{Schekochihin}, A.~A.}, \au{{Numata}, R.} \&
  \au{{Tatsuno}, T.}} \yr{2011}  \at{{Gyrokinetic Simulations of Solar Wind
  Turbulence from Ion to Electron Scales}}.  \jt{Phys. Rev. Lett.}
  \bvol{107}~(3),  \pg{035004}.

\bibitem[{Huang} {\em et~al.\/}(1998){Huang}, {Shen}, {Long}, {Wu}, {Shih},
  {Zheng}, {Yen}, {Tung} \& {Liu}]{1998huang}
{\sc \au{{Huang}, N.~E.}, \au{{Shen}, Z.}, \au{{Long}, S.~R.}, \au{{Wu},
  M.~C.}, \au{{Shih}, H.~H.}, \au{{Zheng}, Q.}, \au{{Yen}, N.~C.}, \au{{Tung},
  C.~C.} \& \au{{Liu}, H.~H.}} \yr{1998}  \at{{The empirical mode decomposition
  and the Hilbert spectrum for nonlinear and non-stationary time series
  analysis}}.  \jt{Proceedings of the Royal Society of London Series A}
  \bvol{454}~(1971),  \pg{903--998}.

\bibitem[{Iroshnikov}(1963)]{1963iroshnikov}
{\sc \au{{Iroshnikov}, P.~S.}} \yr{1963}  \at{{Turbulence of a Conducting Fluid
  in a Strong Magnetic Field}}.  \jt{Astronomicheskii Zhurnal}  \bvol{40},
  \pg{742+}.

\bibitem[{Kida} \& {Ohkitani}(1992)]{1992kida}
{\sc \au{{Kida}, S.} \& \au{{Ohkitani}, K.}} \yr{1992}  \at{{Spatiotemporal
  intermittency and instability of a forced turbulence}}.  \jt{Physics of
  Fluids A}  \bvol{4}~(5),  \pg{1018--1027}.

\bibitem[{Kiyani} {\em et~al.\/}(2015){Kiyani}, {Osman} \&
  {Chapman}]{2015kiyani}
{\sc \au{{Kiyani}, K.~H.}, \au{{Osman}, K.~T.} \& \au{{Chapman}, S.~C.}}
  \yr{2015}  \at{{Dissipation and heating in solar wind turbulence: from the
  macro to the micro and back again}}.  \jt{Philosophical Transactions of the
  Royal Society of London Series A}  \bvol{373}~(2041),
  \pg{20140155--20140155}.

\bibitem[{Koga} {\em et~al.\/}(2007){Koga}, {Chian}, {Miranda} \&
  {Rempel}]{2007koga}
{\sc \au{{Koga}, D.}, \au{{Chian}, A.~C.~L.}, \au{{Miranda}, R.~A.} \&
  \au{{Rempel}, E.~L.}} \yr{2007}  \at{{Intermittent nature of solar wind
  turbulence near the Earth's bow shock: Phase coherence and non-Gaussianity}}.
   \jt{Phys. Rev. E}  \bvol{75}~(4),  \pg{046401}.

\bibitem[{Kraichnan}(1965)]{1965kraichnan}
{\sc \au{{Kraichnan}, R.~H.}} \yr{1965}  \at{{Inertial-Range Spectrum of
  Hydromagnetic Turbulence}}.  \jt{Physics of Fluids}  \bvol{8}~(7),
  \pg{1385--1387}.

\bibitem[{Kuzzay} {\em et~al.\/}(2019){Kuzzay}, {Alexandrova} \&
  {Matteini}]{2019kuzzay}
{\sc \au{{Kuzzay}, D.}, \au{{Alexandrova}, O.} \& \au{{Matteini}, L.}}
  \yr{2019}  \at{{Local approach to the study of energy transfers in
  incompressible magnetohydrodynamic turbulence}}.  \jt{Phys. Rev. E}
  \bvol{99}~(5),  \pg{053202}.

\bibitem[{Landi} {\em et~al.\/}(2015){Landi}, {Del Zanna}, {Papini}, {Pucci} \&
  {Velli}]{2015landi}
{\sc \au{{Landi}, S.}, \au{{Del Zanna}, L.}, \au{{Papini}, E.}, \au{{Pucci},
  F.} \& \au{{Velli}, M.}} \yr{2015}  \at{{Resistive Magnetohydrodynamics
  Simulations of the Ideal Tearing Mode}}.  \jt{Astrophys. J.}  \bvol{806},
  \pg{131}.

\bibitem[{Landi} {\em et~al.\/}(2019){Landi}, {Franci}, {Papini}, {Verdini},
  {Matteini} \& {Hellinger}]{2019landi_arxiv}
{\sc \au{{Landi}, S.}, \au{{Franci}, L.}, \au{{Papini}, E.}, \au{{Verdini},
  A.}, \au{{Matteini}, L.} \& \au{{Hellinger}, P.}} \yr{2019}  \at{{Spectral
  anisotropies and intermittency of plasma turbulence at ion kinetic scales}}.
  \jt{arXiv e-prints}  \pg{p. arXiv:1904.03903},  \arxiv{arXiv: 1904.03903}.

\bibitem[Lin {\em et~al.\/}(2009)Lin, Wang \& Zhou]{2009lin}
{\sc \au{Lin, L.}, \au{Wang, Y.} \& \au{Zhou, H.}} \yr{2009}  \at{Iterative
  filtering as an alternative algorithm for empirical mode decomposition}.
  \jt{Advances in Adaptive Data Analysis}  \bvol{01}~(04),  \pg{543--560},
  \arxiv{arXiv: https://doi.org/10.1142/S179353690900028X}.

\bibitem[{Lion} {\em et~al.\/}(2016){Lion}, {Alexandrova} \&
  {Zaslavsky}]{2016lion}
{\sc \au{{Lion}, S.}, \au{{Alexandrova}, O.} \& \au{{Zaslavsky}, A.}} \yr{2016}
   \at{{Coherent Events and Spectral Shape at Ion Kinetic Scales in the Fast
  Solar Wind Turbulence}}.  \jt{\apj}  \bvol{824}~(1),  \pg{47},  \arxiv{arXiv:
  1602.07213}.

\bibitem[{Loureiro} \& {Boldyrev}(2017)]{2017loureiro}
{\sc \au{{Loureiro}, N.~F.} \& \au{{Boldyrev}, S.}} \yr{2017}
  \at{{Collisionless Reconnection in Magnetohydrodynamic and Kinetic
  Turbulence}}.  \jt{Astrophys. J.}  \bvol{850},  \pg{182}.

\bibitem[{Mallet} {\em et~al.\/}(2017){Mallet}, {Schekochihin} \&
  {Chandran}]{2017mallet}
{\sc \au{{Mallet}, A.}, \au{{Schekochihin}, A.~A.} \& \au{{Chandran},
  B.~D.~G.}} \yr{2017}  \at{{Disruption of Alfv{\'e}nic turbulence by magnetic
  reconnection in a collisionless plasma}}.  \jt{J. Plasma Phys.}
  \bvol{83}~(6),  \pg{905830609}.

\bibitem[{Marsch} \& {Tu}(1997)]{1997marsch}
{\sc \au{{Marsch}, E.} \& \au{{Tu}, C.~Y.}} \yr{1997}  \at{{Intermittency,
  non-Gaussian statistics and fractal scaling of MHD fluctuations in the solar
  wind}}.  \jt{Nonlinear Processes in Geophysics}  \bvol{4}~(2),
  \pg{101--124}.

\bibitem[{Matthews}(1994)]{1994matthews}
{\sc \au{{Matthews}, A.~P.}} \yr{1994}  \at{{Current Advance Method and Cyclic
  Leapfrog for 2D Multispecies Hybrid Plasma Simulations}}.  \jt{J. of Comput.
  Phys.}  \bvol{112},  \pg{102--116}.

\bibitem[{Moffatt} {\em et~al.\/}(1994){Moffatt}, {Kida} \&
  {Ohkitani}]{1994moffatt}
{\sc \au{{Moffatt}, H.~K.}, \au{{Kida}, S.} \& \au{{Ohkitani}, K.}} \yr{1994}
  \at{{Stretched vortices - the sinews of turbulence; large-Reynolds-number
  asymptotics}}.  \jt{Journal of Fluid Mechanics}  \bvol{259},  \pg{241--264}.

\bibitem[{Papini} {\em et~al.\/}(2019{\natexlab{{\em a\/}}}){Papini}, {Franci},
  {Landi}, {Hellinger}, {Verdini} \& {Matteini}]{2019papini_sohe}
{\sc \au{{Papini}, E.}, \au{{Franci}, L.}, \au{{Landi}, S.}, \au{{Hellinger},
  P.}, \au{{Verdini}, A.} \& \au{{Matteini}, L.}} \yr{2019{\natexlab{{\em
  a\/}}}}  \at{{Statistics of magnetic reconnection and turbulence in Hall-MHD
  and hybrid-PIC simulations}}.  \jt{Nuovo Cimento C Geophysics Space Physics
  C}  \bvol{42}~(1),  \pg{23}.

\bibitem[{Papini} {\em et~al.\/}(2019{\natexlab{{\em b\/}}}){Papini}, {Franci},
  {Landi}, {Verdini}, {Matteini} \& {Hellinger}]{2019papini_turb}
{\sc \au{{Papini}, E.}, \au{{Franci}, L.}, \au{{Landi}, S.}, \au{{Verdini},
  A.}, \au{{Matteini}, L.} \& \au{{Hellinger}, P.}} \yr{2019{\natexlab{{\em
  b\/}}}}  \at{{Can Hall Magnetohydrodynamics Explain Plasma Turbulence at
  Sub-ion Scales?}}  \jt{Astrophys. J.}  \bvol{870}~(1),  \pg{52},
  \arxiv{arXiv: 1810.02210}.

\bibitem[{Papini} {\em et~al.\/}(2019{\natexlab{{\em c\/}}}){Papini}, {Landi}
  \& {Del Zanna}]{2019papini_rec}
{\sc \au{{Papini}, E.}, \au{{Landi}, S.} \& \au{{Del Zanna}, L.}}
  \yr{2019{\natexlab{{\em c\/}}}}  \at{{Fast Magnetic Reconnection: Secondary
  Tearing Instability and Role of the Hall Term}}.  \jt{Astrophys. J.}
  \bvol{885}~(1),  \pg{56},  \arxiv{arXiv: 1906.06779}.

\bibitem[Papini {\em et~al.\/}(2018)Papini, Landi \& Zanna]{2018papini}
{\sc \au{Papini, E.}, \au{Landi, S.} \& \au{Zanna, L.~D.}} \yr{2018}  \at{Fast
  magnetic reconnection: The ideal tearing instability in classic, hall, and
  relativistic plasmas.}  \jt{J. Phys.: Conf. Series}  \bvol{1031}~(1),
  \pg{012020}.

\bibitem[{Piersanti} {\em et~al.\/}(2018){Piersanti}, {Materassi}, {Cicone},
  {Spogli}, {Zhou} \& {Ezquer}]{2018piersanti}
{\sc \au{{Piersanti}, M.}, \au{{Materassi}, M.}, \au{{Cicone}, A.},
  \au{{Spogli}, L.}, \au{{Zhou}, H.} \& \au{{Ezquer}, R.~G.}} \yr{2018}
  \at{{Adaptive Local Iterative Filtering: A Promising Technique for the
  Analysis of Nonstationary Signals}}.  \jt{Journal of Geophysical Research
  (Space Physics)}  \bvol{123}~(1),  \pg{1031--1046}.

\bibitem[{Pucci} {\em et~al.\/}(2017){Pucci}, {Servidio}, {Sorriso-Valvo},
  {Olshevsky}, {Matthaeus}, {Malara}, {Goldman}, {Newman} \&
  {Lapenta}]{2017pucci}
{\sc \au{{Pucci}, F.}, \au{{Servidio}, S.}, \au{{Sorriso-Valvo}, L.},
  \au{{Olshevsky}, V.}, \au{{Matthaeus}, W.~H.}, \au{{Malara}, F.},
  \au{{Goldman}, M.~V.}, \au{{Newman}, D.~L.} \& \au{{Lapenta}, G.}} \yr{2017}
  \at{{Properties of Turbulence in the Reconnection Exhaust: Numerical
  Simulations Compared with Observations}}.  \jt{Astrophys. J.}
  \bvol{841}~(1),  \pg{60}.

\bibitem[{Schekochihin} {\em et~al.\/}(2009){Schekochihin}, {Cowley},
  {Dorland}, {Hammett}, {Howes}, {Quataert} \& {Tatsuno}]{2009scheko}
{\sc \au{{Schekochihin}, A.~A.}, \au{{Cowley}, S.~C.}, \au{{Dorland}, W.},
  \au{{Hammett}, G.~W.}, \au{{Howes}, G.~G.}, \au{{Quataert}, E.} \&
  \au{{Tatsuno}, T.}} \yr{2009}  \at{{Astrophysical Gyrokinetics: Kinetic and
  Fluid Turbulent Cascades in Magnetized Weakly Collisional Plasmas}}.
  \jt{Astrophys. J. Suppl.}  \bvol{182}~(1),  \pg{310--377}.

\bibitem[{Servidio} {\em et~al.\/}(2012){Servidio}, {Valentini}, {Califano} \&
  {Veltri}]{2012servidio}
{\sc \au{{Servidio}, S.}, \au{{Valentini}, F.}, \au{{Califano}, F.} \&
  \au{{Veltri}, P.}} \yr{2012}  \at{{Local Kinetic Effects in Two-Dimensional
  Plasma Turbulence}}.  \jt{\prl}  \bvol{108}~(4),  \pg{045001}.

\bibitem[Shay {\em et~al.\/}(2001)Shay, Drake, Rogers \& Denton]{2001shay}
{\sc \au{Shay, M.~A.}, \au{Drake, J.~F.}, \au{Rogers, B.~N.} \& \au{Denton,
  R.~E.}} \yr{2001}  \at{Alfv\'enic collisionless magnetic reconnection and the
  {{Hall}} term}.  \jt{Journal of Geophysical Research}  \bvol{106}~(A3),
  \pg{3759}.

\bibitem[{Spogli} {\em et~al.\/}(2019){Spogli}, {Piersanti}, {Cesaroni},
  {Materassi}, {Cicone}, {Alfonsi}, {Romano} \& {Ezquer}]{2019spogli}
{\sc \au{{Spogli}, L.}, \au{{Piersanti}, M.}, \au{{Cesaroni}, C.},
  \au{{Materassi}, M.}, \au{{Cicone}, A.}, \au{{Alfonsi}, L.}, \au{{Romano},
  V.} \& \au{{Ezquer}, R.~G.}} \yr{2019}  \at{{Role of the external drivers in
  the occurrence of low-latitude ionospheric scintillation revealed by
  multi-scale analysis}}.  \jt{Journal of Space Weather and Space Climate}
  \bvol{9},  \pg{A35}.

\bibitem[Stallone {\em et~al.\/}(2020)Stallone, Cicone, Materassi \&
  Zhou]{2020stallone}
{\sc \au{Stallone, A.}, \au{Cicone, A.}, \au{Materassi, M.} \& \au{Zhou, H.}}
  \yr{2020}  \at{New insights and best practices for the successful use of
  empirical mode decomposition, iterative filtering and derived algorithms}.
  \jt{submitted} .

\bibitem[{Stawarz} {\em et~al.\/}(2016){Stawarz}, {Eriksson}, {Wilder},
  {Ergun}, {Schwartz}, {Pouquet}, {Burch}, {Giles}, {Khotyaintsev}, {Le
  Contel}, {Lindqvist}, {Magnes}, {Pollock}, {Russell}, {Strangeway},
  {Torbert}, {Avanov}, {Dorelli}, {Eastwood}, {Gershman}, {Goodrich},
  {Malaspina}, {Marklund}, {Mirioni} \& {Sturner}]{2016stawarz}
{\sc \au{{Stawarz}, J.~E.}, \au{{Eriksson}, S.}, \au{{Wilder}, F.~D.},
  \au{{Ergun}, R.~E.}, \au{{Schwartz}, S.~J.}, \au{{Pouquet}, A.}, \au{{Burch},
  J.~L.}, \au{{Giles}, B.~L.}, \au{{Khotyaintsev}, Y.}, \au{{Le Contel}, O.},
  \au{{Lindqvist}, P.~A.}, \au{{Magnes}, W.}, \au{{Pollock}, C.~J.},
  \au{{Russell}, C.~T.}, \au{{Strangeway}, R.~J.}, \au{{Torbert}, R.~B.},
  \au{{Avanov}, L.~A.}, \au{{Dorelli}, J.~C.}, \au{{Eastwood}, J.~P.},
  \au{{Gershman}, D.~J.}, \au{{Goodrich}, K.~A.}, \au{{Malaspina}, D.~M.},
  \au{{Marklund}, G.~T.}, \au{{Mirioni}, L.} \& \au{{Sturner}, A.~P.}}
  \yr{2016}  \at{{Observations of turbulence in a Kelvin-Helmholtz event on 8
  September 2015 by the Magnetospheric Multiscale mission}}.  \jt{Journal of
  Geophysical Research (Space Physics)}  \bvol{121}~(11),  \pg{11,021--11,034}.

\bibitem[{Sulem} {\em et~al.\/}(2016){Sulem}, {Passot}, {Laveder} \&
  {Borgogno}]{Sulem_al_2016}
{\sc \au{{Sulem}, P.~L.}, \au{{Passot}, T.}, \au{{Laveder}, D.} \&
  \au{{Borgogno}, D.}} \yr{2016}  \at{{Influence of the Nonlinearity Parameter
  on the Solar Wind Sub-ion Magnetic Energy Spectrum: FLR-Landau Fluid
  Simulations}}.  \jt{Astrophys. J.}  \bvol{818},  \pg{66}.

\bibitem[{Verscharen} {\em et~al.\/}(2019){Verscharen}, {Klein} \&
  {Maruca}]{2019verscharen}
{\sc \au{{Verscharen}, D.}, \au{{Klein}, K.~G.} \& \au{{Maruca}, B.~A.}}
  \yr{2019}  \at{{The multi-scale nature of the solar wind}}.  \jt{Living
  Reviews in Solar Physics}  \bvol{16}~(1),  \pg{5}.

\bibitem[{Wan} {\em et~al.\/}(2012){Wan}, {Matthaeus}, {Karimabadi},
  {Roytershteyn}, {Shay}, {Wu}, {Daughton}, {Loring} \& {Chapman}]{2012wan}
{\sc \au{{Wan}, M.}, \au{{Matthaeus}, W.~H.}, \au{{Karimabadi}, H.},
  \au{{Roytershteyn}, V.}, \au{{Shay}, M.}, \au{{Wu}, P.}, \au{{Daughton}, W.},
  \au{{Loring}, B.} \& \au{{Chapman}, S.~C.}} \yr{2012}  \at{{Intermittent
  Dissipation at Kinetic Scales in Collisionless Plasma Turbulence}}.
  \jt{Phys. Rev. Lett.}  \bvol{109}~(19),  \pg{195001}.

\bibitem[{Wan} {\em et~al.\/}(2015){Wan}, {Matthaeus}, {Roytershteyn},
  {Karimabadi}, {Parashar}, {Wu} \& {Shay}]{2015wan}
{\sc \au{{Wan}, M.}, \au{{Matthaeus}, W.~H.}, \au{{Roytershteyn}, V.},
  \au{{Karimabadi}, H.}, \au{{Parashar}, T.}, \au{{Wu}, P.} \& \au{{Shay}, M.}}
  \yr{2015}  \at{{Intermittent Dissipation and Heating in 3D Kinetic Plasma
  Turbulence}}.  \jt{Phys. Rev. Lett.}  \bvol{114}~(17),  \pg{175002}.

\bibitem[{Widmer} {\em et~al.\/}(2016){Widmer}, {B{\"u}chner} \&
  {Yokoi}]{2016widmer}
{\sc \au{{Widmer}, F.}, \au{{B{\"u}chner}, J.} \& \au{{Yokoi}, N.}} \yr{2016}
  \at{{Sub-grid-scale description of turbulent magnetic reconnection in
  magnetohydrodynamics}}.  \jt{Physics of Plasmas}  \bvol{23}~(4),
  \pg{042311},  \arxiv{arXiv: 1511.04347}.

\bibitem[{Wu} {\em et~al.\/}(2013){Wu}, {Perri}, {Osman}, {Wan}, {Matthaeus},
  {Shay}, {Goldstein}, {Karimabadi} \& {Chapman}]{2013wu}
{\sc \au{{Wu}, P.}, \au{{Perri}, S.}, \au{{Osman}, K.}, \au{{Wan}, M.},
  \au{{Matthaeus}, W.~H.}, \au{{Shay}, M.~A.}, \au{{Goldstein}, M.~L.},
  \au{{Karimabadi}, H.} \& \au{{Chapman}, S.}} \yr{2013}  \at{{Intermittent
  Heating in Solar Wind and Kinetic Simulations}}.  \jt{Astrophys. J. Lett.}
  \bvol{763}~(2),  \pg{L30}.

\bibitem[Wu \& Huang(2009)]{2009wu}
{\sc \au{Wu, Z.} \& \au{Huang, N.~E.}} \yr{2009}  \at{Ensemble empirical mode
  decomposition: a noise-assisted data analysis method}.  \jt{Advances in
  Adaptive Data Analysis}  \bvol{01}~(01),  \pg{1--41},  \arxiv{arXiv:
  https://doi.org/10.1142/S1793536909000047}.

\bibitem[{Yang} {\em et~al.\/}(2017){Yang}, {Matthaeus}, {Parashar},
  {Haggerty}, {Roytershteyn}, {Daughton}, {Wan}, {Shi} \& {Chen}]{2017yang}
{\sc \au{{Yang}, Y.}, \au{{Matthaeus}, W.~H.}, \au{{Parashar}, T.~N.},
  \au{{Haggerty}, C.~C.}, \au{{Roytershteyn}, V.}, \au{{Daughton}, W.},
  \au{{Wan}, M.}, \au{{Shi}, Y.} \& \au{{Chen}, S.}} \yr{2017}  \at{{Energy
  transfer, pressure tensor, and heating of kinetic plasma}}.  \jt{Physics of
  Plasmas}  \bvol{24}~(7),  \pg{072306}.

\end{thebibliography}

\end{document}